\documentclass[amsmath,preprintnumbers,10pt,twocolumn,longbibliograpy,prl]{revtex4-1}
\usepackage{amsbsy}
\usepackage{graphicx}
\usepackage{color}
\usepackage{subfigure}
\usepackage{footnote}
\usepackage{bm}
\usepackage{color}
\usepackage{xr}
\usepackage{graphicx}
\usepackage{xkeyval,xcolor}
\makeatletter
\newlength{\sfp@hseplen}\newlength{\sfp@vseplen}
\define@cmdkey{subfigpos}[sfp@]{pos}[ul]{}
\define@cmdkey{subfigpos}[sfp@]{font}[\small]{}
\define@cmdkey{subfigpos}[sfp@]{vsep}[1.5\baselineskip]{\setlength{\sfp@vseplen}{\sfp@vsep}}
\define@cmdkey{subfigpos}[sfp@]{hsep}[7pt]{\setlength{\sfp@hseplen}{\sfp@hsep}}
\newcommand{\subfigimg}[3][,]{%
  \setkeys{Gin,subfigpos}{pos,font,vsep,hsep,#1}
  \setbox1=\hbox{\includegraphics{#3}}
  \ifnum\pdfstrcmp{\sfp@pos}{ul}=0
    \leavevmode\rlap{\usebox1}
    \rlap{\hspace*{\sfp@hsep}\raisebox{\dimexpr\ht1-\sfp@vsep}{\sfp@font{#2}}}
    \phantom{\usebox1}
  \else\ifnum\pdfstrcmp{\sfp@pos}{ur}=0
    \leavevmode\usebox1
    \llap{\raisebox{\dimexpr\ht1-\sfp@vsep}{\sfp@font{#2}}\hspace*{\sfp@hsep}}
  \else\ifnum\pdfstrcmp{\sfp@pos}{lr}=0
    \leavevmode\usebox1
    \llap{\raisebox{\sfp@vsep}{\sfp@font{#2}}\hspace*{\sfp@hsep}}
  \else
    \leavevmode\rlap{\usebox1}
    \rlap{\hspace*{\sfp@hseplen}\raisebox{\sfp@vsep}{\sfp@font{#2}}}
    \phantom{\usebox1}
  \fi\fi\fi
}
\makeatother

\begin{document}
\title{Energy dissipation and fluctuations in a driven liquid}
\author{Clara del Junco, Laura Tociu, Suriyanarayanan Vaikuntanathan}
\affiliation{Department of Chemistry and The James Franck Institute, University of Chicago, Chicago, IL, 60637}

\begin{abstract}
Minimal models of active and driven particles have recently been used to elucidate many properties non-equilibrium systems. However, the relation between energy consumption and changes in the structure and transport properties of these non-equilibrium materials remains to be explored. We explore this relation in a minimal model of a driven liquid that settles into a time periodic steady state. Using concepts from stochastic thermodynamics and liquid state theories, we show how the work performed on the system by the various non-conservative, time dependent forces \textendash this quantifies a violation of time reversal symmetry \textendash modifies the structural, transport, and phase transition properties of the driven liquid.
\end{abstract}

\maketitle
\section{Introduction}

Minimal models of active matter have provided an analytically and computationally tractable test bed to study non-equilibrium systems. 
Phase transitions in some classes of model systems composed of self-propelled particles are beginning to be characterized~\cite{Tailleur2008,Speck2014,Speck2015,Cates2015,Takatori2015}. Recent work has also studied nucleation phenomena~\cite{Richard2016,Redner2016} and obtained expressions for pressure and other mechanical properties of active media~\cite{Solon2015a,Solon2015c,Speck2016a,Bialke2015}.  
In spite of these advances, understanding the connections between energy consumption and the structural properties of these systems remains a challenging problem~\cite{Battle604}.

In our work, we explore these connections in a model non-equilibrium liquid driven by time-periodic forces. The rotational dynamics that result from the driving have similarities with a range of systems, including colloids in a periodically changing magnetic or electric field~\cite{Yan2015, Han2016,Wysocki2009,Vissers2011a}, shaken plastic particles and chiral wires~\cite{Grunwald2016,Tsai2005}, and chemical and biological microswimmers with active rotational degrees of freedom~\cite{Kummel2013,Maggi2015,Lauga2006,Riedel2005,Drescher2009,Petroff2015}. 
Yet despite their relevance to this wide range of experimental and biological systems,  model systems with rotational dynamics have only recently begun to be studied~\cite{VanZuiden2016,Nguyen2014b,Dasbiswas2017}.

The central points of this paper are as follows: First, we describe the class of driven liquids with rotational dynamics considered in this paper and their associated phase diagrams. Second, we show that the density fluctuations in our non-equilibrium system are surprisingly well described by Gaussian statistics~\cite{Chandler1993}. Within this effective description, we are able to derive simple scaling relations for the work performed on the system due to the non-equilibrium driving forces. These analytical predictions are validated by simulation data from the full many-particle systems. Third, we derive a relation that shows how the rate of work done on the model system changes the fluctuations in the conservative forces experienced by the particles. In other words, our relation describes how energy dissipation changes the structural properties of the non-equilibrium material. This relation between work and force fluctuations can be viewed as an instantiation of the Harada-Sasa relation~\cite{Harada2005,Harada2006} that connects work performed in a non-equilibrium system to a breakdown of the fluctuation dissipation relation. We demonstrate this relation numerically for a variety of driven systems to firmly establish its general nature. Finally, using a minimal model we demonstrate how the aforementioned breakdown of the fluctuation dissipation relation can change the diffusion constant of the system. While our driven liquid is substantially more complex than the minimal model, this provides intuition for the interplay between dissipation and transport. Crucially, it provides a model for how the phase transition properties of the driven liquid depend on the magnitude of the driving force. Taken together, our results elucidate how a violation of time reversal symmetry can be used to alter the structural, transport, and phase transition properties of a liquid.
\section{Phase behavior of a driven system}
\begin{figure}[h!]
\centering
\subfigure{
\includegraphics[width=0.8\linewidth]{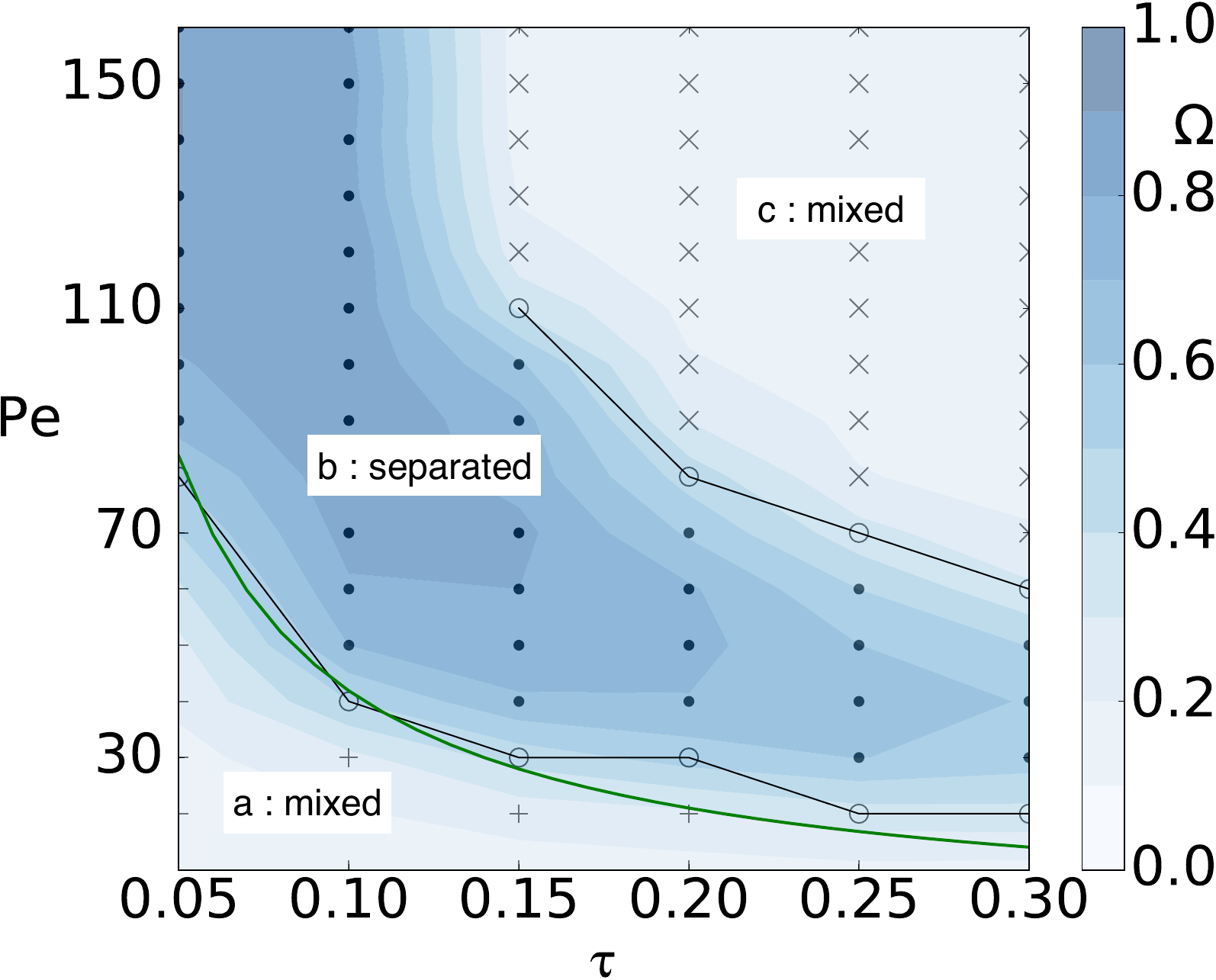}}
\subfigure{
\subfigimg[width=0.28\linewidth, pos=ul, font=\large\textsf]{a}{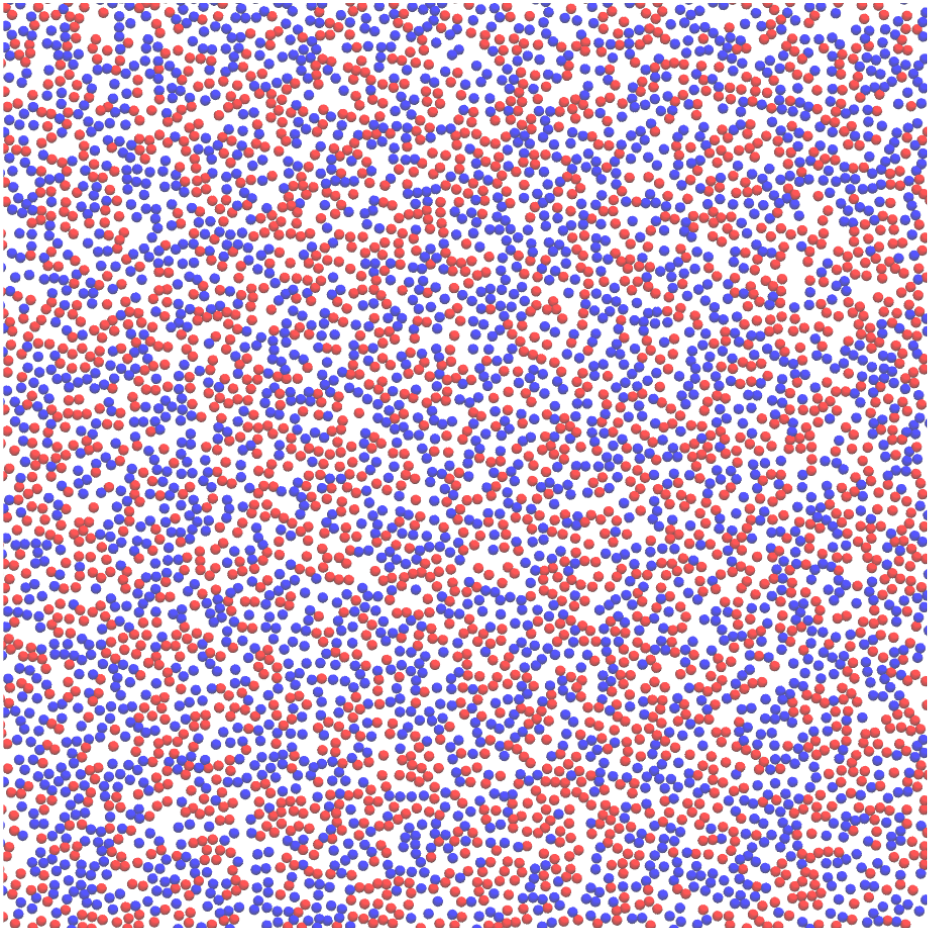}}
\subfigure{
\subfigimg[width=0.28\linewidth, pos=ul, font=\large\textsf]{b}{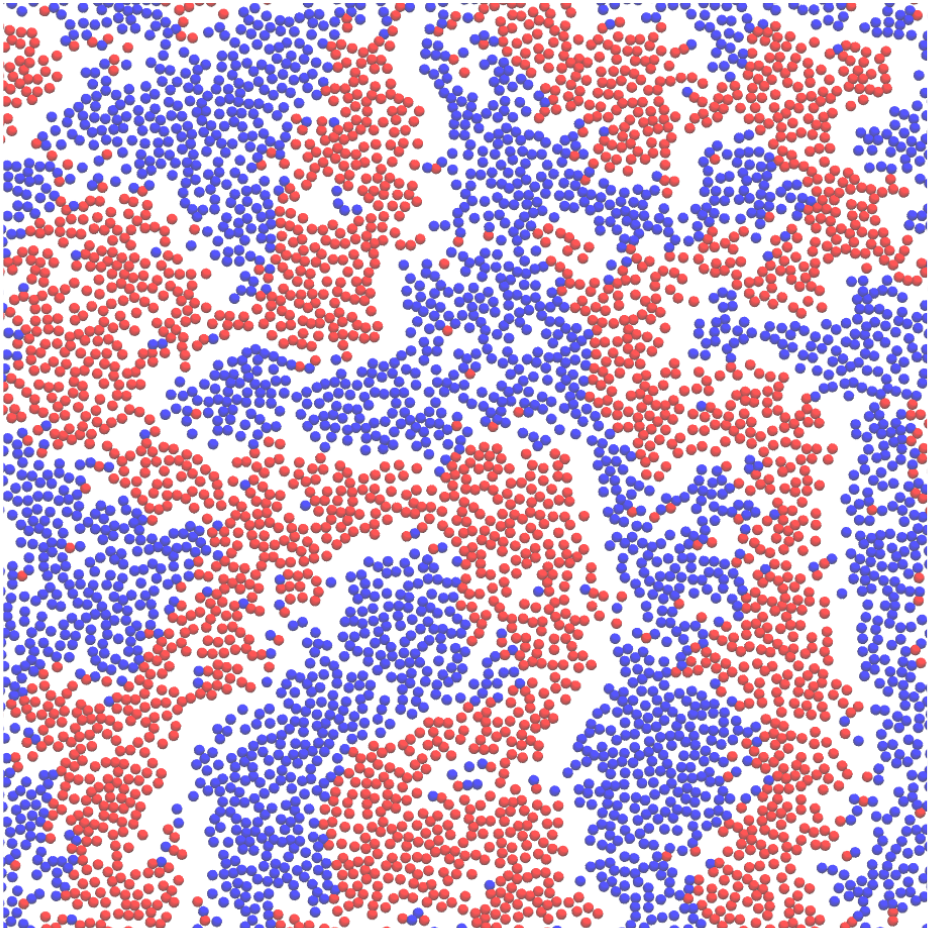}}
\subfigure{
\subfigimg[width=0.28\linewidth, pos=ul, font=\large\textsf]{c}{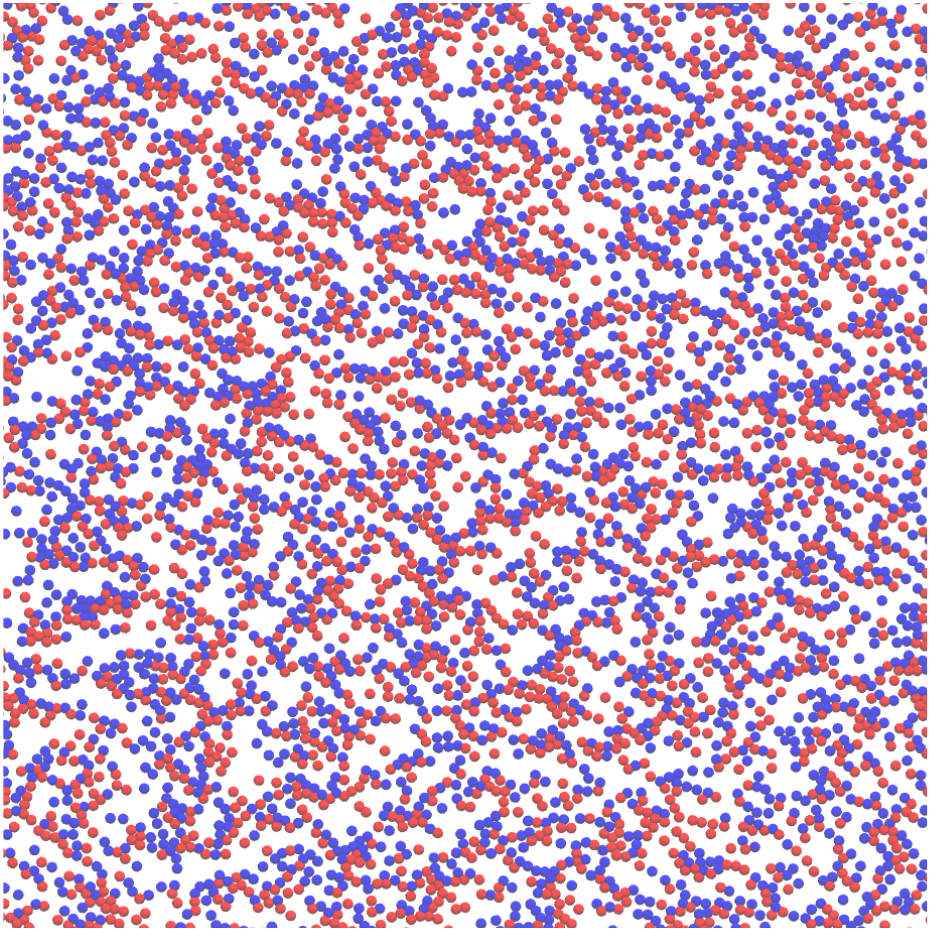}}
\caption{(Top) Phase diagram of the system with circular driving at number density $\rho=0.45$. At low $Pe$, the system is mixed (plus signs). As $Pe$ is increased, the system undergoes a transition to a phase separated state (points). As $Pe$ is increased further, there is a transition to a mixed state characterized by large variations in local density (crosses). Letters refer to representative snapshots below. The locations of the transitions are defined as the maxima of $\partial \Omega/\partial Pe$ (black lines). The green line is proportional to the work as given by Eq.~\ref{eq:workScaling}, where we have used the mean values of $\gamma_G$ and $K_G$ found from the fits in Fig.~\ref{fig:WvsPeandTau}.  Contours were drawn using interpolated values. (Bottom) Snapshots of the final frame of a simulation in different regions of the phase diagram at $\rho=0.45$. (a) At $Pe=15,\tau=0.1$, (b) At $Pe=70,\tau=0.15$, (c) At $Pe=150,\tau=0.2$. Driven particles are colored red, and undriven particles are colored blue.}
\label{fig:PhaseDiagram}
\end{figure}
Our model system is composed of purely repulsive 2-dimensional disks whose positions evolve in time according to driven Brownian dynamics:
\begin{equation}
{\bf{\dot r}_i}(t) = D_0\beta\left({\bf F}_{\rm c,i}(t) + {\bf F}_{\rm d}(t)\right)+\boldsymbol\eta_i (t)
\label{eq:xEOM}
\end{equation}
where $D_0$ is the bare diffusion constant, ${\bf F}_{\rm c,i}$ is given by the derivative of the Weeks-Chandler-Anderson interaction potential~\cite{WCA1971}, and $\boldsymbol\eta_i (t) = (\eta_{i,x}(t),\eta_{i,y}(t))$  are Gaussian-distributed random variables with $\left<\boldsymbol\eta_i (t)\right>=0$ and $\left<\eta_{i,\mu}(t)\eta_{j,\nu}(t')\right> = 2D_0\delta_{i,j}\delta_{\mu,\nu}\delta(t-t')$. In all of our simulations and calculations, we set $\beta=1/k_BT=1$. All results reported here are for a number density of $\rho=N/L^2=0.45$. In addition to the conservative forces, half of the particles are driven by an external force acting on the center of mass of the particle whose direction changes periodically in time according to:
\begin{align}
&{\bf F}_{\rm d}=A\sin\theta \hat e_x+A\cos\theta\hat e_y \label{eq:FexEq} \\
&\theta=2\pi t/\tau. \label{eq:theta}
\end{align}
For the other half of the particles, ${\bf F}_{\rm d}=0$. Thus, in the zero-temperature limit, a single driven particle will trace a circle in the plane. ${\bf F}_{\rm d}$ is the same for all driven particles, so that a pure system of driven particles, if we move to a frame of reference that is rotating with ${\bf F}_{\rm d}$, will look the same as the equilibrium system. This model was motivated in part by a recent experimental active matter system developed by Luijten, Granick and coworkers~\cite{Han2016}.

The driving force in Eq.~\ref{eq:FexEq} is characterized by a period $\tau$ and an amplitude $A$. The latter is quantified by the Peclet number, a dimensionless measure of the ratio of advective to diffusive velocity in the system. Here we define it as $Pe = \frac{A/\gamma}{D_0/r_0}$, where $A$ is the amplitude of the driving force. The bare, single-particle diffusion coefficient ($D_0$), particle radius ($r_0$), and friction coefficient ($\gamma$) are all set to 1 in our simulations. The period of the driving force is measured in units of time set by $t_0=r_0^2/D_0$.
We found 3 distinct phases in the region of $Pe,\tau$ space that we studied here (Fig.\,\ref{fig:PhaseDiagram}). We characterized the phases using an order parameter $\Omega$ (defined in Methods) that measures the degree of mixing. Briefly, a small value of $\Omega$ indicates a mixed system while a large value of $\Omega$ indicates a demixed system.
At low $Pe$, driven and passive particles remain mixed and the system is homogeneous. As $Pe$ is increased, they segregate into regions of purely driven or passive particles. Similar to other strongly damped active systems with rotating dynamics, the interfaces have no particular orientation, and there is a particle current along the interface that decays rapidly in to the bulk~\cite{Sabrina2015}. The steady state is time-periodic with a period $\tau$. As $Pe$ is increased further, the system undergoes a transition to a mixed phase characterized by large variations in the local particle number density. Through both transitions, the value of $\Omega$ changes smoothly. As $\tau$ is increased, phase separation persists over a smaller range of $Pe$. The curve separating the regions $a$ and $b$ in the phase diagram in Fig.~\ref{fig:PhaseDiagram} (and in phase diagrams generated with other ratios of $A_x/A_y$) are well described by the functional form $Pe^2 f(\tau)= c$, where $c$ is a constant.

We now  investigate  how the non-equilibrium forces modify the structural and transport properties, in particular the diffusion constant,  of this driven liquid. In order to study how these properties change as we approach phase separation, we focus here on the lower mixed region of phase space (Snapshot a, Fig.~\ref{fig:PhaseDiagram}).
We then use the relation between diffusion and energy input to explain the dependence of the phase transition between regions $a$ and $b$ in Fig.~\ref{fig:PhaseDiagram} on $Pe$.

\section{Gaussian density fluctuations in the driven liquid}

In the context of equilibrium liquids, integral equation theories~\cite{MTOC} can be used to demonstrate that, within certain approximations~\cite{Dean1996,Chandler1993,Kruger2017}, the coarse-grained field $\delta\rho ({\bf r},t)= \rho({\bf r},t)-\bar\rho$, where $\bar\rho$ is the bulk density of the liquid, can satisfy the following equation of motion:
\begin{equation}
\dot{\delta\rho} ({\bf r},t) = -\frac{K_G}{\gamma_G}\delta\rho({\bf r},t)+\eta_G({\bf r},t).
\label{eq:deltarhoEOM}
\end{equation}
Here, $K_G$ is an effective spring constant related to the derivative of a Hamiltonian $\delta H/\delta (\delta\rho)$, $\gamma_G$ is an effective friction, and $\eta_G$ is an effective noise with statistics $\left<\eta_G(t)\right>=0$ and $\left<\eta_G(t)\eta_G(t')\right> = 2\delta(t-t')\delta({\bf r}-{\bf r'})D_G$. We provide an expression for $K_G$ in the SI (SI Sec.\thinspace II B ). 

Eq.~\ref{eq:deltarhoEOM} implies that the density fluctuations effectively behave as a particle diffusing in a harmonic potential, and that the probability distribution of $\delta\rho$ values are Gaussian. In many equilibrium liquids ranging from simple hard spheres to water, the statistics of fluctuations in the number of particles inside a small probe volume of the system are Gaussian~\cite{Chandler1993,Hummer1996}. This property justifies Eq.~\ref{eq:deltarhoEOM} and has enabled the development of quantitatively accurate theories for the thermodynamics of these liquids~\cite{Vaikuntanathan2014b,Vaikuntanathan2016necessity,Hummer1996}. 

\begin{figure}
\center
\includegraphics[angle=0, width=0.7\linewidth]{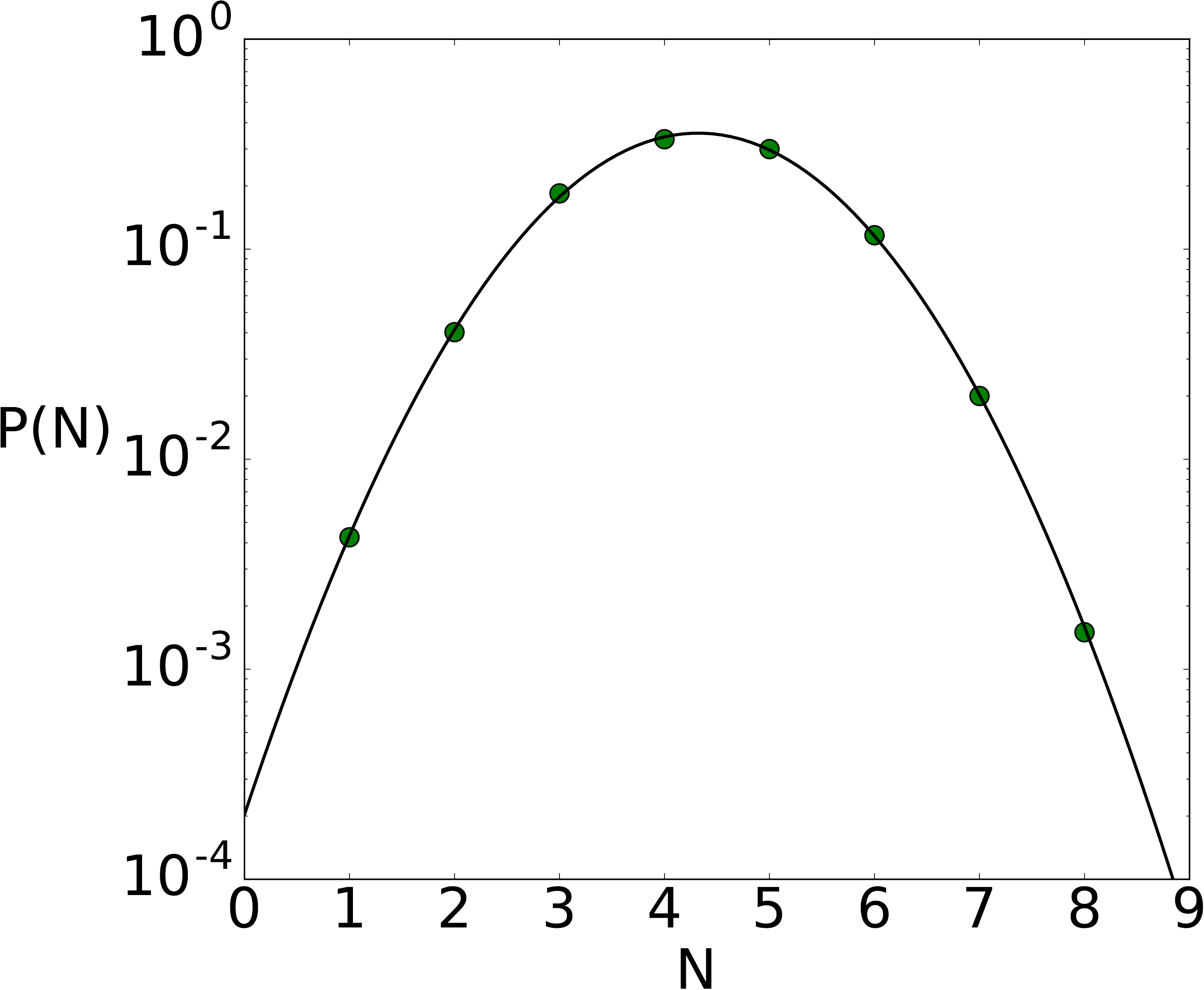}
\caption{Number density probability distribution for a system with $Pe=20, \tau=0.1$. Points are the measured probability of finding N particles in a randomly selected probe volume of radius $1.75r_0$. The line is a Gaussian distribution with mean $\left<N\right> = 4.32$ and variance $\sigma^2 = 1.25$. Error bars are smaller than the symbols. The density statistics are Gaussian to approximately $\pm3\sigma$ from the mean. Number density statistics are similarly Gaussian for all of the points represented in Fig.~\ref{fig:WvsPeandTau}-\ref{fig:WvsD}.}
\label{fig:Gaussian}
\end{figure}

To determine whether a Gaussian theory can describe the fluctuations out of equilibrium, we measured the statistics of number density fluctuations in the driven liquid. We find that they are indeed Gaussian to a good approximation at points on the phase diagram that are well below the line where the system first phase separates. One example histogram is shown in Fig.~\ref{fig:Gaussian}. 

This finding is unexpected for a couple of reasons. First, it is surprising even in equilibrium liquids that density fluctuations in small volumes would be Gaussian. Assuming that the process of particles entering and exiting some small region of the system is Markovian, with a defined entry and exit rate, the naive expectation is that the resulting probability distribution would be a Poisson distribution. It is only in the limit of very large probe volumes that the distribution would tend to be Gaussian, as a result of the central limit theorem. Yet, as noted, Gaussian statistics in small volumes of many equilibrium liquids are well-documented~\cite{Chandler1993,Hummer1996}. Second, intuition and experience give us little reason to expect that the statistics would remain Gaussian out of equilibrium -- for example, `giant' (i.e. non-Gaussian) number fluctuations have been observed in some active matter systems\cite{Narayan2007,McCandlish2012,Palacci2013}. 

\begin{figure}
\center
\subfigure{
\includegraphics[width=0.7\linewidth]{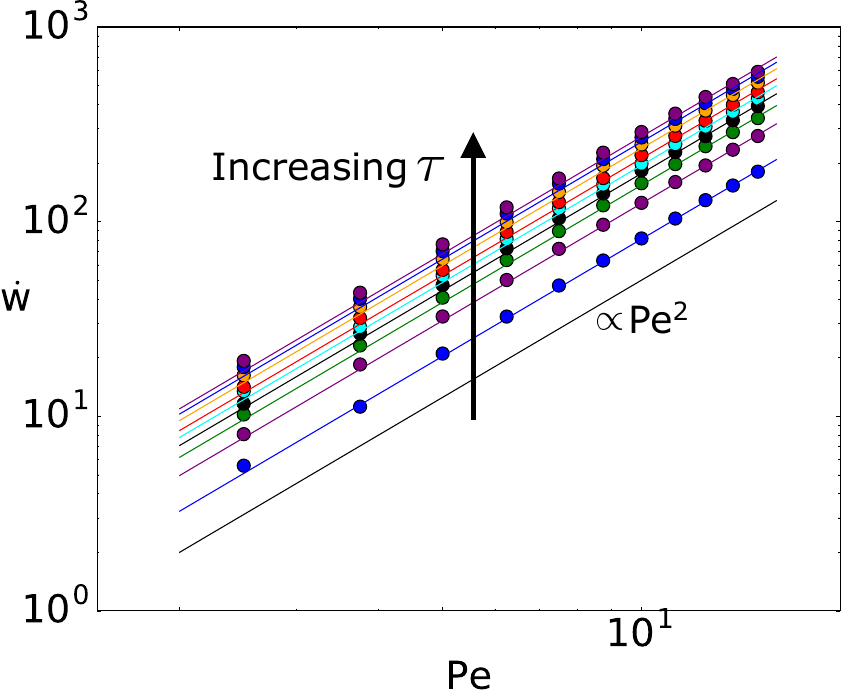}}
\subfigure{
\includegraphics[width=0.7\linewidth]{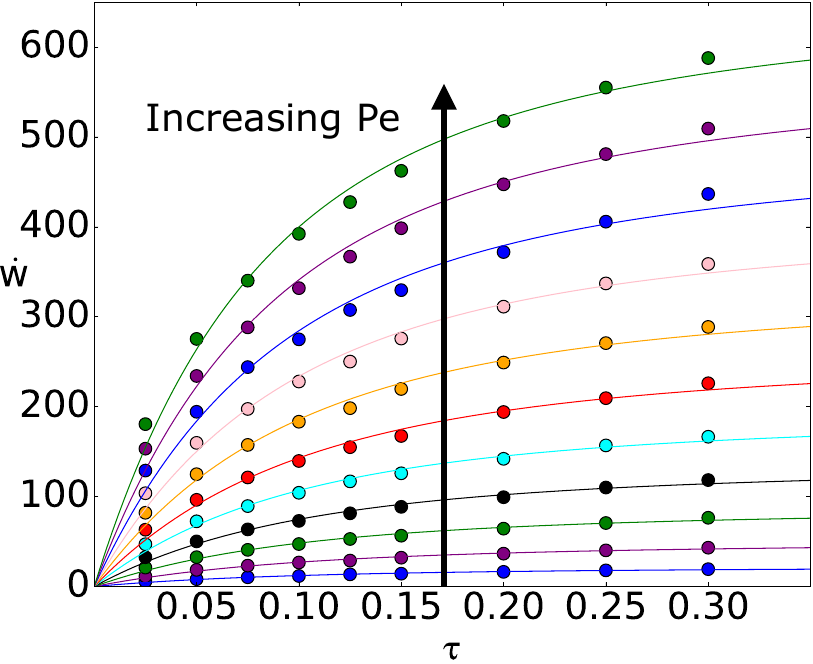}}
\caption{Rate of work done on the system as a function of $Pe$ (top) and $\tau$ (bottom) by the driving forces in the full many-particle simulation, for $Pe$ ranging from 2.5-15 and $\tau$ ranging from 2.5-30. Lines are fits to Eq.~\ref{eq:workScaling}. Parameters $\gamma_G$ and $K_G$ were fit separately to each line; the variation in $\gamma_G$ is less than 20\% between curves; the variation in $K_G$ is less than 5\%. Error bars are smaller than the points.}
\label{fig:WvsPeandTau}
\end{figure}

The Gaussian nature of density fluctuations also allows us to predict how the work performed on the non-equilibrium liquid scales with $Pe$ and $\tau$. In order to extract this scaling we add an extra driving term $F_{G,d}$ to Eq.~\ref{eq:deltarhoEOM}:
\begin{equation}
\dot{\delta\rho} ({\bf r},t) = -\frac{K_G}{\gamma_G}\delta\rho({\bf r},t)+\frac{F_{G,d}}{\gamma_G}+\eta_G({\bf r},t).
\label{eq:deltarhoEOMdriven}
\end{equation}
If $F_{G,d}=0$, we recover the equilibrium limit Eq.~\ref{eq:deltarhoEOM}. To mimic the local increase in density as an active particle is driven into surrounding passive particles, we imagine a process where the average density is changed linearly at a rate $Pe$, for a duration $\tau$. This driving protocol corresponds to $F_{G,d} = (K_GPe) t$. We emphasize that we have not derived this driving term from the microscopic equations of motion -- rather, we are postulating a minimal protocol based on the physical picture. The average rate of work performed during this Gaussian process can be computed analytically, and is equal to~\cite{Horowitz2009}:
\begin{equation}
\label{eq:workScaling}
\left<\dot w\right>=\gamma_G Pe^2\left[ 1 -\frac{\gamma_G}{\tau K_G}\left[1-e^{-K_G \tau/\gamma_G}\right]\right].
\end{equation}
In Fig.~\ref{fig:WvsPeandTau}, we show that $\dot w$ in the atomistic simulations does indeed scale as predicted in Eq.~\ref{eq:workScaling}, with fitting parameters $\gamma_G$ and $K_G$. We have checked that the scaling also holds for different driving protocols including cases where the particles are not phase locked or have random phases, and in systems with unequal number fraction of driven and undriven particles (SI Fig.\thinspace 4). These results strongly suggest that the scaling of the work with $Pe$ and $\tau$ are consequences of the Gaussian nature of density fluctuations.  

\section{Dissipation modifies the statistics of force fluctuations}
\begin{figure}
\center
\includegraphics[width=0.7\linewidth]{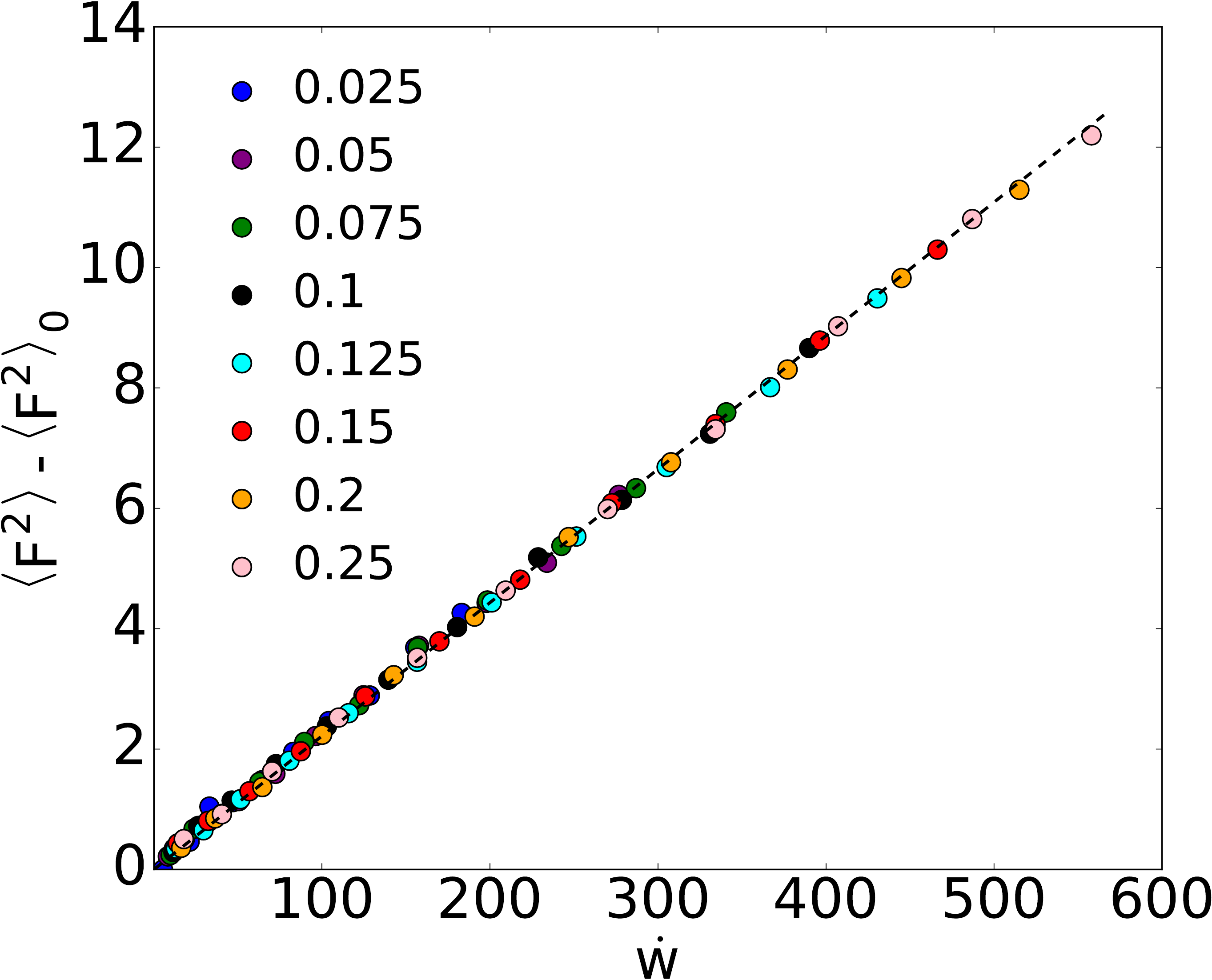}
\caption{Change in the force correlations as a function of the rate of work done on the system. Colors indicate values of $\tau$. The data collapse on to the line predicted by the Gaussian form in Eq.~\ref{eq:wForces}. Error bars are smaller than the points.}
\label{fig:WvsForces}
\end{figure}

We now  demonstrate how the work done on the system by the non-equilibrium forces affects its microscopic statistics.  We begin by noting that according to Floquet theory~\cite{Hanggi1998}, the non-equilibrium steady state induced by time-periodic driving forces is also time-periodic. For the system to achieve such a time-periodic steady state, the increase in the internal energy of the system due to the total work, $w$, over each cycle has to be dissipated as heat.

The rate of heat emitted by the driven system (per particle) can be conveniently expressed in terms of the following stochastic integral,
\begin{equation}
\label{eq:qDef}
\dot {q}  =-\frac{1}{N} \sum_{i=1}^{N}\int {\bf F}_i \cdot \dot{{\bf r}_i}
\end{equation}
where ${\bf F}_i$ and ${\bf r}_i$ denote the conservative force and position vectors for particle $i$, $\dot{{\bf r}_i}$ is the rate of change of the position vector of particle $i$ due to forces other than the non-conservative force, the sum is over all the particles in the system, and the integral is performed in the Stratonovich sense~\cite{Harada2006,Sekimoto1997}. The average rate of heat emission $\langle \dot{q}\rangle$ should equal the average rate at which work is performed on the system, $\langle \dot{w} \rangle$. 

Before commenting on the effect of dissipative heat and work fluxes in the  many body microscopic system, we first apply the  definition in Eq.~\ref{eq:qDef} to the simpler Gaussian picture in Eq.~\ref{eq:deltarhoEOMdriven} above. Specifically, as we show in SI Sec.\thinspace I, the average heat flux of a non-equilibrium system with Gaussian modes can be written as 
\begin{equation}
-\left<\dot q \right>_G=\frac{1}{\gamma_G} \left[ \left<F_G^2\right> - \left<F_G^2\right>_0 \right]= \left<\dot w \right>_G
\label{eq:wForces}
\end{equation}
where $\gamma_G$ and $F_G=-K_G \delta\rho$ are the effective friction coefficient and conservative force for the Gaussian modes, respectively. The second term in Eq.~\ref{eq:wForces} $\left(\left<F_G^2\right>_0/\gamma_G\right)$ is defined as $-\frac{1}{2}\lim_{t \to 0}\left<\eta_G(0)F_G(t)\right>$.  The $\left<F_G^2\right>$ term describes the spontaneous fluctuations of the particle in the well, while the $\left<F_G^2\right>_0$ term describes the relaxation of the system following a random perturbation~\cite{Harada2005}. In  SI Sec.\thinspace II A,  we show that in equilibrium these two terms are equal (a consequence of fluctuation dissipation theorem) and no work is done on the system on average. Out of equilibrium, the work performed is positive and the relation between fluctuations and response breaks down. In this case, the difference between fluctuations and response is predicted exactly by flux of heat, or alternately the rate at which work is performed, as illustrated in Eq.~\ref{eq:wForces}. We note that this expression is completely insensitive to the driving protocol; that is, we can use any $F_{G,d}$ we want in Eq.~\ref{eq:deltarhoEOMdriven}.

The forces on the RHS in Eq.~\ref{eq:wForces} are generalized forces that act on the density fluctuations, and \textit{not} the forces that appear in the individual particle equations of motion. It is reasonable to ask whether a relation similar to Eq.~\ref{eq:wForces} in terms of ${\bf F}$, the sum of pair-wise conservative forces acting on a tagged particle in the fluid, could exist. We find that it does.  At equilibrium, we show in SI Sec.\thinspace III that the fluctuations $\left<{\bf F}^2\right>$ exactly predict the response $\left<{\bf F}^2\right>_0$ (SI Sec.\thinspace III). Moreover, far from equilibrium, we find in Fig.~\ref{fig:WvsForces} that the work performed in the many-particle system is indeed related to the breakdown of the equilibrium relation between fluctuations and response,
\begin{equation}
\label{eq:wForcesatomistic}
\left<\dot w \right>\propto \left<{\bf F}^2\right> - \left<{\bf F}^2\right>_0 \, .
\end{equation}
Similar to the interpretation in Eq.~\ref{eq:wForces}, $\left<{\bf F}^2\right>_0$ captures the response of the non-equilibrium liquid following a random perturbation while $\left<{\bf F}^2\right>$ describes the force fluctuations on a tagged particle in its non-equilibrium steady state.

This relation is an instantiation of the Harada-Sasa relation~\cite{Harada2005,Harada2006} and can be obtained from stochastic thermodynamics arguments similar to those used for the Gaussian model. Briefly, the equation of motion for an undriven particle in our liquid is given by Eq.~\ref{eq:xEOM} with ${\bf F}_{d}(t)=0$. Multiplying both sides of this equation by ${\bf F}_{c,i}(t)$, interpreting the resultant equation in the Stratonovich sense~\cite{Harada2005}, and using Eq.~\ref{eq:qDef}, we obtain the atomistic equivalent of Eq.~\ref{eq:wForces} connecting the heat flux to statistics of force fluctuations. Since the heat and work fluxes have to balance each other in steady state to ensure a constant average energy, we obtain Eq.~\ref{eq:wForcesatomistic}. 
We verified Eq.~\ref{eq:wForcesatomistic} for a wide range of amplitudes and time periods of the driving force in the homogeneous part of the phase diagram. We have also verified that this result holds for driving forces in which the particles are not phase locked, have random phases, and in systems with unequal number fraction of driven and undriven particles (SI Fig.\thinspace 4). The relation also holds when applied separately to driven and undriven particles in all the cases -- in other words, the work performed on average due to the driven particles predicts the change in the force fluctuations of the undriven particles. 

We now study the implications of this result for the diffusion constant of the non-equilibrium system.

\section{Enhanced diffusion due to non-conservative forces}

\begin{figure}
\center
\includegraphics[width=0.7\linewidth]{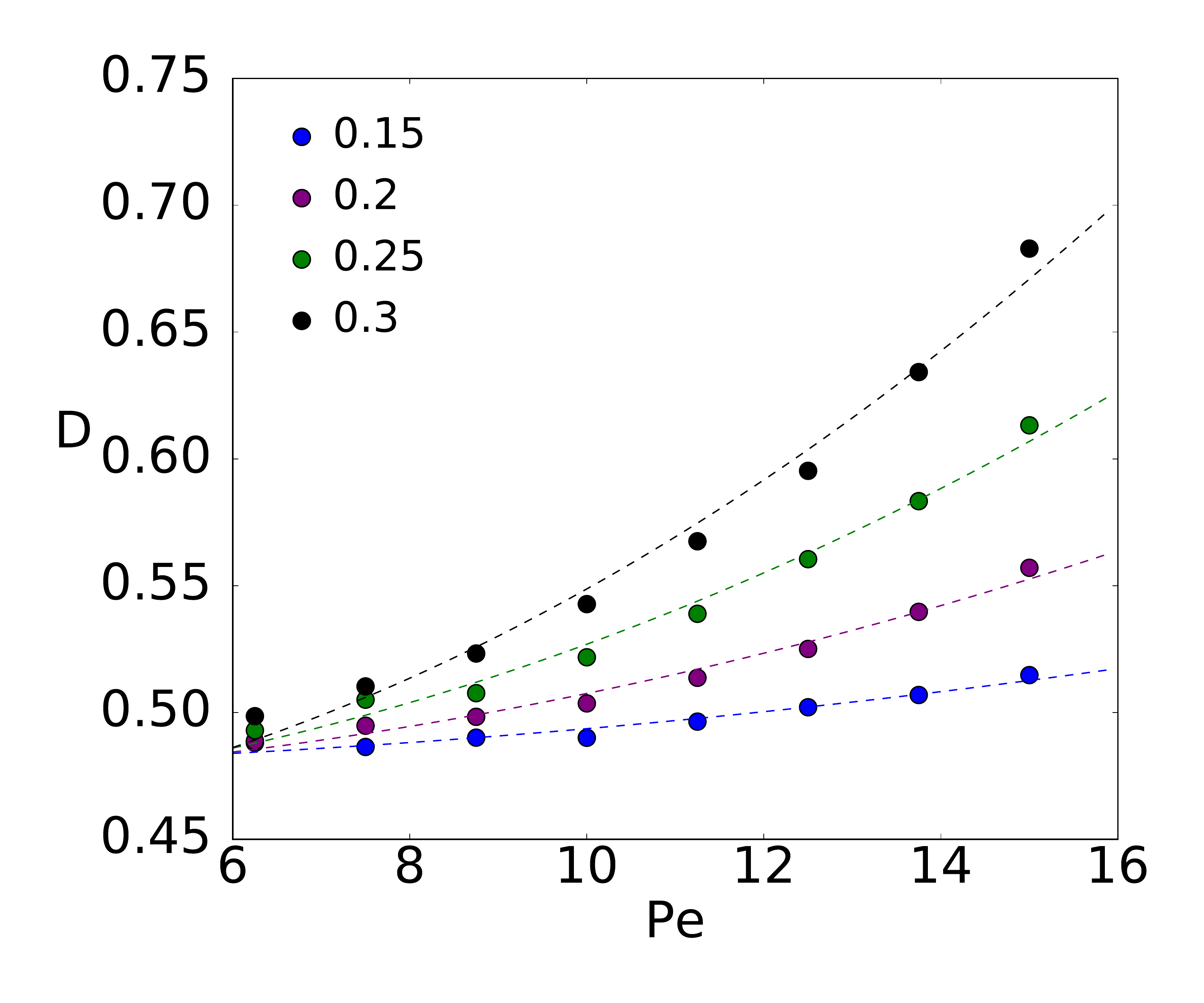}
\caption{Scaling of the diffusion coefficient of a tagged particle in the liquid, $D$, with the amplitude of the driving force, $Pe$. Colors indicate values of $\tau$. Error bars are smaller than the points. The renormalization of the diffusion constant, $D-D_{eq}$, is well predicted by the quadratic relation $D-D_{eq}\propto Pe^2$.}
\label{fig:WvsD}
\end{figure}

Since the diffusion is related as follows to force autocorrelation functions in Brownian dynamics, 
\begin{equation}
D-D_0= \frac{1}{d} \int_0^\infty  \left[\left<\frac{{\bf F}(0) \cdot {\bf F}(t)}{\gamma^2}\right>+\left<\frac{{\boldsymbol \eta}(0) \cdot {\bf F}(t)}{\gamma}\right>\right]{d}t \,,
\label{eq:KGD1}
\end{equation}
where $d$ denotes the dimensionality of the system, it is reasonable to ask whether the change in the force correlations that accompanies breakdown of the fluctuation dissipation relation, Eq.~\ref{eq:wForcesatomistic}, affects the diffusion coefficient of the driven liquid.
From a physical point of view we may expect the diffusion constant to increase with the driving forces.  The driving forces do work on the system by inducing collisions between driven and undriven particles (but not between particles of the same type); these collisions scatter the particles, which can increase the diffusion coefficient. 

Indeed, from simulations we find that the diffusion constant increases with the amplitude of the driving force, $Pe$. This increase is well approximated by $D-D_0\propto Pe^2$ (Fig.~\ref{fig:WvsD}). Understanding the basis of the increase of $D$ due to the nonconservative forces is important -- in the present context, it can help explain how the phase transition depends on the non-equilibrium forces. We note that non-conservative forces need not necessarily lead to an increase in the diffusion constant: in the self-propelled variety of active matter systems, the diffusion constant eventually decreases as a function of the driving force, leading to self-trapping and phase transitions\cite{Burkholder2017,Hancock2017}. 

In order to understand the dependence of $D$ on the non-conservative forces, we consider a minimal model of a tracer particle diffusing in a fluid
\begin{equation}
\dot{{\bf r}}=h {\bf \tilde F}({\bf r}) +\boldsymbol{\tilde{\eta}}(t),
\label{eq:model}
\end{equation}
where ${\bf \tilde F}({\bf r})$ is spatially dependent force that can be designed to model the forces acting on a tagged particle in the liquid, $h$ is a parameter that tunes the coupling between the fluid and the tracer particle, and $\boldsymbol{\tilde{\eta}}(t)$ is a Gaussian delta function correlated white noise. In order to ensure no net drift, we constrain $\int  {\bf \tilde F} ({\bf r}) d{\bf r} =0$. We imagine sampling over many realizations of the force ${\bf \tilde F}({\bf r})$ from a distribution in order to model the forces exerted by the fluid on a tracer particle. In the liquid considered in the previous sections, the statistics of force fluctuations on a tagged particle satisfy $\langle {\bf F}^2 \rangle=\langle {\bf F}^2\rangle_0$ when the system is in equilibrium. Equilibrium dynamics in Eq.~\ref{eq:model} are achieved whenever ${\bf \tilde F}({\bf r})$ can be expressed as a gradient of a  potential, ${\bf \tilde F}({\bf r})=-\nabla \tilde U(\bf {r})$. In such cases, it can be demonstrated that the equivalent relation $\langle \tilde {\bf F}^2 \rangle=\langle \tilde {\bf F}^2\rangle_0$ holds, where the averages $\langle\dots\rangle$ are taken both over the statistics of the random noise $\boldsymbol{\tilde{\eta}}(t)$ and over many realizations of the force. The system can be driven out of equilibrium by ensuring that the force in Eq.~\ref{eq:model} has a non-conservative component, ${\bf \tilde F}({\bf r})= {\bf \tilde F}_c({\bf r}) + Pe {\bf \tilde F}_d({\bf r})$, where ${\bf \tilde F}_c ({\bf r}) = -\nabla \tilde U({\bf r})$, ${\bf \tilde F}_d (\bf r) = \nabla \times {\bf \tilde A}({\bf r})$ and we choose ${\bf \tilde A}({\bf r})$ such that $\nabla \cdot {\bf \tilde A}({\bf r}) = 0$. Like in the previous sections, $Pe$ tunes the magnitude of the non-conservative force, ${\bf \tilde F}_d$. In this case, similar to the many-particle driven liquid, a breakdown of the fluctuation dissipation relation is predicted by the total amount of entropy dissipated by the system~\cite{Harada2006}. Specifically, we use a perturbation theory~\cite{Demery2014,Fisher1984} (described in the SI Sec.\thinspace IV, V, VI) to show that to $O(h^2)$,
\begin{equation}
\langle \dot\sigma\rangle = \langle \tilde {\bf F}^2 \rangle-\langle \tilde {\bf F}^2\rangle_0= \frac{Pe^2 h^2}{V} \int  \langle {\bf \tilde F}_{d}({\bf q}) \cdot {\bf \tilde F}_{d}(-{\bf q})\rangle d{\bf q}
\end{equation}
where $\langle \dot\sigma\rangle$ denotes the average rate of entropy dissipation, analogous to $\langle \dot{w}\rangle$ in the many-body liquid.  

For this minimal model, we obtained expressions for the diffusion constant in Eq.~\ref{eq:model} to $O(h^2)$:

\begin{equation}
\begin{aligned}
D-D_0 =&\alpha \frac{h^2}{V}\left(Pe^2 \int  \langle |{\bf \tilde A}({\bf r})|^2 \rangle d{\bf r} -  \int \langle \tilde U^2({\bf r})\rangle d{\bf r} \right)  
\label{eq:diffusion1}
\end{aligned}
\end{equation}
where $\alpha=D_0/(d (k_BT)^2)$, $d$ is the dimension, and we set $\langle \tilde U \rangle=0$ without loss of generality so that $\langle \tilde U^2\rangle$ is simply the variance of energy fluctuations. 
In instances where the spectrum of force fluctuations is strongly peaked at a particular wave vector ${\bf q}^*$, the diffusion constant can be simply related to the average entropy dissipation rate:

\begin{equation}
D-D_0= \alpha \frac{\langle \dot{\sigma}\rangle}{|{\bf q}^*|^2}- \alpha h^2 \langle \tilde U^2\rangle.
\label{eq:diffusion2}
\end{equation}

The dynamics of our driven liquid, specified by Eq.~\ref{eq:xEOM}, are substantially more complicated that the minimal model considered. Nonetheless, the expressions in Eq.~\ref{eq:diffusion1} and Eq.~\ref{eq:diffusion2} provide useful insight. By associating the variance  $\langle \tilde U^2\rangle $ with the sampled variance of energy fluctuations of a tagged particle in the many-body driven liquid, and the entropy production rate $\langle \dot{\sigma}\rangle$  with the work performed $\langle \dot{w} \rangle $ in the many-body liquid, Eq.~\ref{eq:diffusion2} demonstrates how the non-conservative forces can modify the diffusion properties of a particle in the fluid. In particular, the minimal model predicts that the diffusion constant can increase as $Pe^2$ in the presence of non-conservative forces and decrease with increasing variance of energy fluctuations. In other words, the diffusion constant need not scale monotonically as $Pe^2$ with driving. Rather, Eq.~\ref{eq:diffusion1} and Eq.~\ref{eq:diffusion2} suggest that the increase of the diffusion constant is bounded by terms that scale as $Pe^2$. The observed scaling from simulation data in Fig.~\ref{fig:WvsD} is consistent with this expectation. 

Finally, we find that the curve separating the low drive mixed phase from the phase separated region in the phase diagram in Fig.~\ref{fig:PhaseDiagram} (and in phase diagrams generated with other ratios of $A_x/A_y$) are well described by the functional form $Pe^2 f(\tau)= c$, where $c$ is a constant.  This feature is consistent with the relation between diffusion and $Pe$ predicted by Eq.~\ref{eq:diffusion2}. Since work is only performed in regions where there is a mixture of driven and undriven particles, Eq.~\ref{eq:diffusion2} suggests that the diffusion constant can inherit the dependence of the work, $w$, on $Pe$ and the composition $f$.  

Note that we are not able to comment on the dependence of $D$ on $\tau$ because our minimal model, Eq.~\ref{eq:model}, doesn't resolve this variable. Rather, the minimal model effectively assumes that $\tau$ is much larger than any timescales inherent to the dynamics of the system. A simple linear stability analysis reveals that such a composition dependent diffusion, $D(f)$, can render the homogeneous phase in region $a$ of Fig.~\ref{fig:PhaseDiagram} unstable with $Pe^2$ determining the location of the transition. 
In this manner, our results explain how the renormalization of the diffusion constant due to the non-equilibrium forces can eventually drive the phase separation.

\section{Discussion \& Conclusions}

From the rich physics of non-equilibrium materials, and in particular of active particles with rotating dynamics, it is clear that dissipation plays an important role in modifying the structural and dynamical properties of the steady states of these systems. Here, in the context of a class of systems with rotating dynamics, we have identified how the rate of work done by the external forces renormalizes force fluctuations. Using simplified descriptions of density fluctuations, we were also able to model how the work performed in this many body system depends on the non-equilibrium forces. Finally, using a minimal model, we explained the observed enhancement of the diffusion due to the non-equilibrium driving forces and proposed a relation between diffusion and dissipation. The analysis based on the minimal model also helped explain the observed dependence of the phase behavior on the magnitude $Pe$ of the driving force. These results demonstrate how the material properties of driven liquids, which are experimentally realizable as colloids, can be tuned simply by violating time reversal symmetry and controlling the amount of energy put in to the system. 

\section{Acknowledgments}
We gratefully acknowledge very useful discussions with D.C., Etienne Fodor, Tom Witten, and Grant Rotskoff. The authors acknowledge support from  NSF DMR-MRSEC 1420709, University of Chicago and the Army Research Office under grant number W911NF-16-1-0415. 

\section{Methods}

{\textit{Model--}}
The system was simulated using the equations of motion given in Eqs.~\ref{eq:xEOM}-\ref{eq:theta}. ${\bf F}_{\rm c,i}$ is the force on particle $i$ which results from a purely repulsive Weeks-Chandler-Anderson interaction potential~\cite{WCA1971}:
\begin{equation} 
u(r_{ij})=\left\{ \begin{matrix} 4\epsilon \left[ \left( \frac{r_0}{r_{ij}} \right)^{12}-\left( \frac{r_0}{r_{ij}}\right)^{6}\right]+\epsilon, & r \leq 2^{1/6}r_0 \\ 
0, & r>2^{1/6}r_0 \end{matrix} \right.
\label{WCAeq}
\end{equation}
where $r_0$  is the particle radius and $\epsilon$ is the interaction strength. The time scale is set by $t_0 = r_0^2/D_0$. All results reported are for a driven particle fraction $f=0.5$ and number density $\rho=N/L^2=0.45$. LAMMPS~\cite{Plimpton1995} was used to integrate the equations of motion.  \\
%
%
{\textit{Phase Diagram--}}
To construct the phase diagram, square simulation boxes of size $50r_0 \times 50r_0$ with periodic boundary conditions were filled with 1125 particles with random initial positions to get $\rho=0.45$. We scanned values of $Pe$ from 10-160 and $\tau$ from 0.05 to 0.3$t_0$. Simulations were run with timestep $\Delta t = 10^{-6} t_0$ up to $t=1000t_0$.
To identify different phases in the system, we defined an order parameter $\Omega$ which measures the degree of segregation of driven and passive particles. To compute $\Omega$, we randomly selected a number of small probe areas with radius $1.75r_0$ in the steady-state system and computed the normalized difference in the local density of driven and passive particles, and then averaged over many locations and many frames.  
For a completely mixed system $\Omega$ is 0, for a completely separated system $\Omega$ is slightly lower than 1 due to the presence of interfaces. We chose a probe volume of radius $1.75r_0$, which is large enough for $\Omega$ to take on a range of values, but small enough that inhomogeneities are not all washed out. Since the value of $\Omega$ is sensitive to the size of probe volume and the overall density of the system, we defined a phase transition as the location where $\partial\Omega/\partial Pe$ at constant $\tau$ is steepest. \\
{\textit{Density Fluctuations--}}
The number of particles $N$ in a randomly selected probe volume of radius $1.75r_0$ was measured in 20 probe volumes per frame in 200 frames of a simulation after the system had reached the steady state. Individual data were binned to construct the histogram.\\
{\textit{Work--}}
We calculated the work by integrating $\dot w=\left<{\bf F}_{\rm c,i} \delta {\bf r}_{\rm i}\right>$, where the $\delta {\bf r}_{\rm i}$ is the displacement of particle $i$ due to the non-conservative forces only in 1 timestep $\Delta t = 10^{-5} t_0$ and ${\bf F}_{\rm c,i}$ is the conservative force. The average is over all particles in the system. Measurements were made for 100$\tau$ after the system reached the steady state. \\
{\textit{Diffusion--}}
Square simulation boxes of size $200r_0 \times 200r_0$ with PBC were filled with 18000 particles. Trajectories were run for at least 1000$\tau$ with a timestep of $\Delta t = 10^{-5} t_0$. The mean-squared-displacement (MSD) of all particles in the simulations was measured after each cycle $\tau$, so that there is no net displacement due to the driving forces.  The diffusion coefficient was extracted by fitting the MSD as a function of time according to $\left<x^2(t)\right> = 4Dt$. \\

%



\end{document}


\title{Energy dissipation and fluctuations in a driven liquid}
\author{Clara del Junco, Laura Tociu, Suriyanarayanan Vaikuntanathan} 
\affiliation{The University of Chicago, Chicago, IL, 60637}

\maketitle

\section{Relation between $\dot q$, $\dot w$ and the forces in the driven system}\label{sec:wdotDerivation}

We postulate the following EOM for the driven Gaussian system:
\begin{equation}
\frac{\partial\delta\rho}{\partial t} = \frac{-K_G}{\gamma_G}(\delta\rho-u(t)) + \eta_G = \frac{F_{G,c} + F_{G,d}}{\gamma_G} + \eta_G \label{eq:deltarhodriven}
\end{equation}

In this section all $F_c,F_{d}, K, \gamma, D, \eta$ shall be understood to be $F_{G,c},F_{G,d}, K_G, \gamma_G, D_G,\eta_G$, the effective conservative force, driving force, spring constant, friction coefficient, diffusion coefficient and noise of the Gaussian density fluctuations, respectively.

We wish to capture the process in which particles are being pushed into each other by the driving forces, thereby changing the average density. This is represented by a time-dependent minimum of the potential given by $u(t)$. We make no assumption here about the form of $u(t)$. To determine the amount of work done on the system, we divide the dynamics in to 2 parts: first, the system is taken from $\delta\rho(t)$ to $\delta\rho'(t)$ by the non-conservative forces only, and then it is taken from $\delta\rho'(t)$ to $\delta\rho(t+\Delta t)$ by the conservative and random forces. In the first step, work is done; no heat is exchanged. In the second step, no work is done; heat is dissipated to the surroundings. If the total energy of the system remains the same, as it does once the system has reached the time-periodic steady state, any work done in the first step must be dissipated as heat in the second step.

The amount of heat dissipated by the system to the bath in the second step is
\begin{equation}
dq= E(t+\Delta t)-E'(t)= - F_c d(\delta\rho).\label{eq:dq}
\end{equation}

We have 2 options for expanding the stochastic integrand on the RHS: the Stratonovich integral or the It\^o integral, and we will use the former. To see why, consider the result of Eq.~\ref{eq:dq} for an equilibrium system if we expand in the It\^o sense:
\begin{equation}
F_c d(\delta\rho) = F_c(t) (\delta\rho(t+\Delta t)-\delta\rho(t)) = \left(\frac{K}{\gamma}\delta\rho(t)\right)\left(\frac{K}{\gamma}\delta\rho(t)+\eta\right)\Delta t
\end{equation}
Since terms that are linear in $\eta$ vanish upon averaging, this gives:
\begin{equation}
\left<F_c d(\delta\rho)\right>= \left(\frac{K}{\gamma}\right)^2\left<\delta\rho^2\right>\Delta t > 0.
\end{equation} 
Clearly, this cannot be true for an equilibrium system since the net heat flow must be zero. To obtain the correct result it is necessary to interpret the integral in the Stratonovich sense:
\begin{equation}
F_c d(\delta\rho) = \frac{\left(F_c(t)+F_c(t+\Delta t)\right)}{2} (\delta\rho(t+\Delta t)-\delta\rho(t)) \propto \delta\rho(t+\Delta t)^2 - \delta\rho(t)^2\, .
\end{equation}
Since the system is at equilibrium, $\left<\delta\rho(t+\Delta t)^2\right> = \left<\delta\rho(t)^2\right>$ and the heat flow is 0 as required. 

Returning to Eq.~\ref{eq:dq}, we now assume that the system is out of equilibrium as a result of the driving force. We have:
\begin{equation}
F_c d(\delta\rho) = \frac{(F_c(t)+F_c(t+\Delta t))}{2} (\delta\rho(t+\Delta t)-\delta\rho'(t)) = -\frac{K}{2} (\delta\rho(t+\Delta t)^2 - \delta\rho'(t)^2),
\end{equation}
but now we can no longer assume that $\left<\delta\rho(t+\Delta t)^2\right> = \left<\delta\rho'(t)^2\right>$. Expanding $\delta\rho(t+\Delta t)$ gives:
\begin{align}
\delta\rho(t+\Delta t)^2 - \delta\rho'(t)^2 = \left[\delta\rho'(t)+\left(-\frac{K}{\gamma}\delta\rho'(t)+\eta\right)\Delta t\right]^2 - \delta\rho'(t)^2 
\end{align}
Assuming a vanishing time step, we only keep terms to first order in $\Delta t$. Terms that are linear in $\eta$ vanish upon averaging. This leaves:
\begin{equation}
\langle dq \rangle=  \frac{K}{2} \left(-\frac{2K\left<\delta\rho'(t)^2\right>\Delta t}{\gamma} + \left<\eta^2\right>\Delta t^2\right)
\end{equation}
By definition, $\left<\eta^2\right>\Delta t^2=2D\Delta t$. Since $\delta\rho'(t) = \delta\rho(t)+u(t)\Delta t/\gamma$, expanding  $\left<\delta\rho'(t) \right>^2\Delta t $ gives $\left<\delta\rho(t) \right>^2\Delta t + O(\Delta t^2) $; so, to first order in $\Delta t$ we can simply replace $\delta\rho'(t)$ with $\delta\rho(t)$ and the form of $u(t)$ is irrelevant. We are left with:
\begin{equation}
\left<dq\right> =  -\frac{K}{2} \left(\frac{2K\left<\delta\rho(t)^2 \right>}{\gamma} - 2D\right)\Delta t \,.
\end{equation}
Finally, we define (see section~\ref{sec:FlucRelax})
\begin{equation}
\left<F^2\right>_0/\gamma \equiv KD \,.
\end{equation}
In the next section, we show that  $\left<F^2\right>_0 / \gamma = -\frac{1}{2}\lim_{t\to 0}\left<\eta(0)F(t)\right>$ (see Eq.~\ref{eq:etaFequals2KD}), so that $\langle F^2\rangle_0/\gamma$ quantifies the response of the system to random perturbations. These considerations give us:
\begin{equation}
\left<\frac{dq}{dt}\right> = -\frac{1}{\gamma}\left ( \left<F^2\right> - \left<F^2\right>_0\right)
\end{equation}
and since $\langle \dot w \rangle = -\langle \dot q\rangle$:
\begin{equation}
\left<\frac{dw}{dt}\right> = \frac{1}{\gamma}\left ( \left<F^2\right> - \left<F^2\right>_0\right).
\end{equation}

\section{Gaussian density fluctuations}\label{sec:Gaussian}

\subsection{Fluctuations and relaxation in equilibrium}\label{sec:FlucRelax}
We first show that $-2\left<F_G(0)F_G(t)\right>_0/\gamma = \left<\eta_G(0)F_G(t)\right>$ in equilibrium. The equation of motion for Gaussian density fluctuations is:

\begin{equation}
\frac{\partial\delta\rho}{\partial t} = \frac{-K_G}{\gamma_G}\delta\rho + \eta_G = \frac{F_G}{\gamma_G} + \eta_G \label{eq:deltarho}
\end{equation}

In the following, all $K, \gamma, D,F, \eta$ shall be understood to be $K_G, \gamma_G, D_G, F_G$, and $\eta_G$, the spring constant, friction coefficient, diffusion coefficient, force, and noise of the Gaussian density fluctuations according to Eq.~\ref{eq:deltarho}, respectively.

The solution of Eq.~\ref{eq:deltarho} is:
\begin{equation}
\delta\rho(t) = \delta\rho(0) e^{-Kt/\gamma}+\int_0^t ds e^{-K(t-s)/\gamma}\eta(s).
\end{equation}

The force-noise correlation is given by:
\begin{align}
\langle \eta(0)F(t)\rangle &= -K\left<\eta(0)\delta\rho(t)\right> \\
&= -K\int_0^t ds e^{-K(t-s)/\gamma}\left<\eta(0)\eta(s)\right>\\
&= -K\int_0^t ds e^{-K(t-s)/\gamma}\delta(s)2D\\
&= -2KDe^{-Kt/\gamma}
\label{eq:etaFequals2KD}
\end{align}

For the Gaussian equation of motion, $\left<\delta\rho^2\right>=1/(\beta K)$, where $\beta=1/(D \gamma)$.  So:

\begin{equation}
KD = K^2\frac{D}{K}=\frac{(K)^2}{\gamma}\frac{1}{\beta K} = \frac{(K)^2\left<\delta\rho^2\right>}{\gamma}=\frac{\left<F^2\right>}{\gamma}
\label{eq:KDequalF2}
\end{equation}

\begin{figure}
\centering
\includegraphics[width=0.5\linewidth]{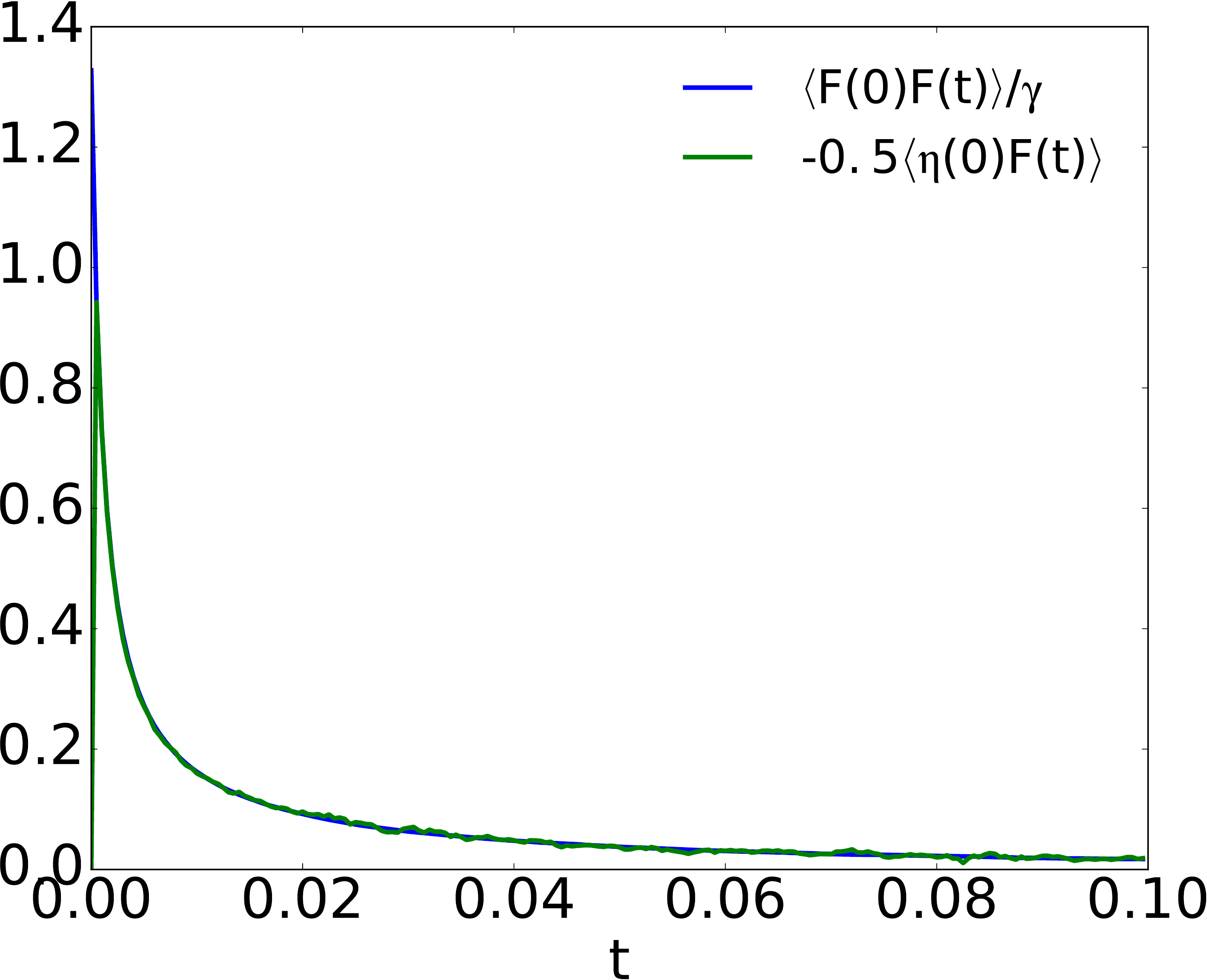}
\caption{Force-force and noise-force correlations in equilibrium. $-0.5\left<\eta(0)F(t)\right>$ is equal to  $\left<F(0)F(t)\right>/\gamma$ at all times. Only times up to $0.1t_0$ are shown here in order to clearly show the short-time behavior. The divergence at $t=0$ is a result of the finite time step used in simulations.}
\label{fig:eqCorrs}
\end{figure}

The force-force correlation is given by:
\begin{align}
\langle F(0)F(t) \rangle &= K^2\left<\delta\rho(0)\delta\rho(t)\right> \\
&= K^2\langle \delta\rho(0)^2\rangle e^{-Kt/\gamma} \\
&= \langle F^2\rangle e^{-Kt/\gamma}
\label{eq:F2equalsKD}
\end{align}
Comparing Eq.~\ref{eq:etaFequals2KD},~\ref{eq:F2equalsKD}, and~\ref{eq:KDequalF2} we find that 
\begin{equation}
-2\left<F_G(0)F_G(t)\right>_0/\gamma = \left<\eta_G(0)F_G(t)\right>
\label{eq:F2equalsetaF}
\end{equation}
for a system that obeys Eq.~\ref{eq:deltarho}.  In Fig.~\ref{fig:eqCorrs}, we show that $-2\left<{\bf F}(0){\bf F}(t)\right>_0/\gamma = \left<{\boldsymbol \eta}(0){\bf F}(t)\right>$ in the many-particle system at equilibrium, where ${\bf F}$ is the total force on a tagged particle due to the WCA potential, ${\boldsymbol \eta}$ is the random force on a tagged particle, and $\gamma$ is the friction coefficient appearing in the individual particle EOM. 

\begin{figure}
\centering
\includegraphics[width=0.5\linewidth]{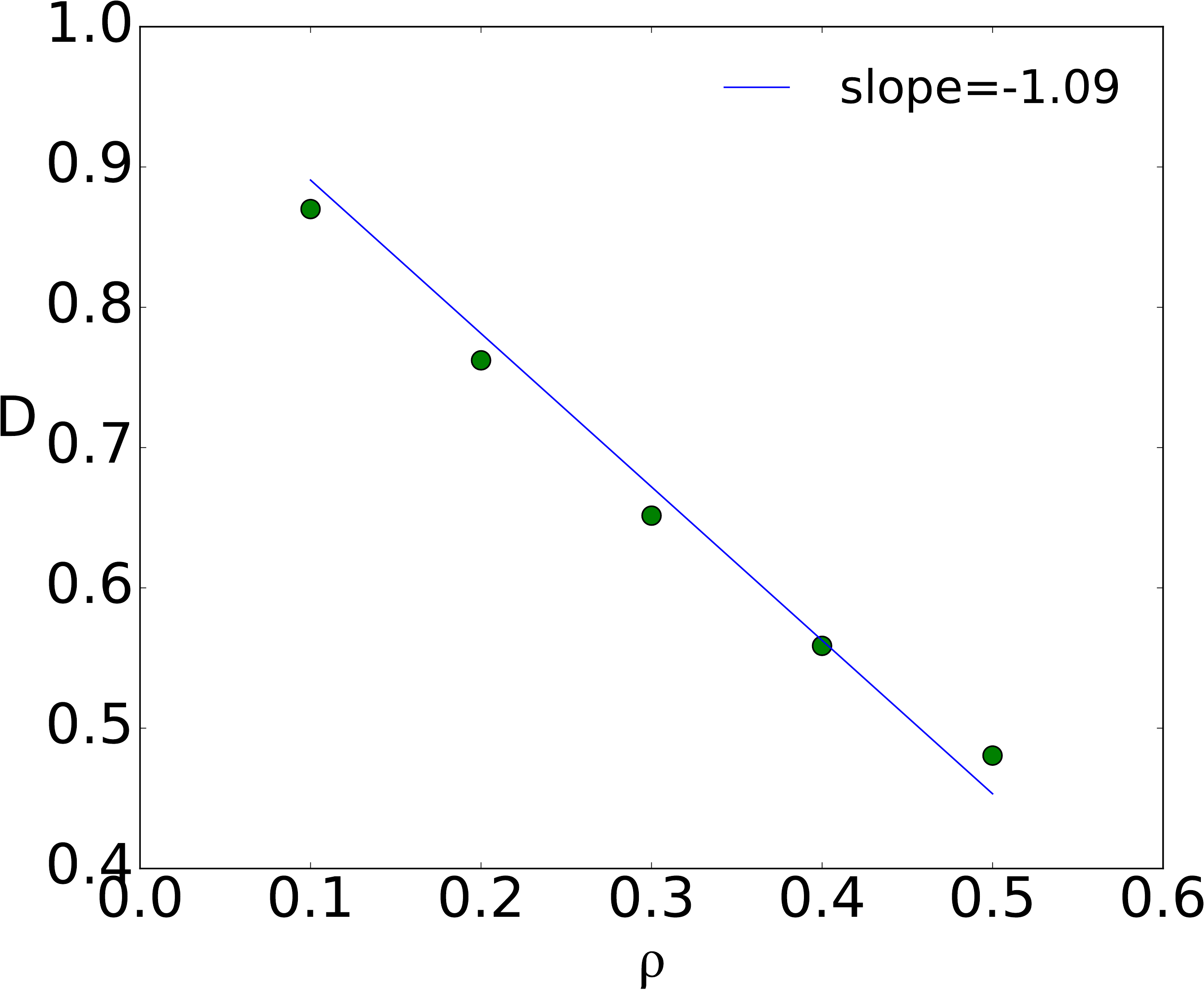}
\caption{The diffusion in the many-particle system at equilibrium as a function of particle number density. The bare diffusion coefficient is $D_0=1$.}
\label{fig:Dvsrho}
\end{figure}

\subsection{Expression for the effective spring constants in the Gaussian representation and connection to atomistic simulations}\label{SpringConstants}

In this section, we provide expressions for the effective spring constant $K_G$ using ideas from liquid state theory.

We begin by writing down a Hamiltonian appropriate for Gaussian density fluctuations in a liquid~\cite{Chandler1993}:
\begin{equation}
H^G = \frac{\beta}{2} \int d \mathbf r \int d \mathbf r' \delta\rho(\mathbf r) \chi^{-1}(\mathbf r,\mathbf r') \delta\rho(\mathbf r)
\end{equation}
$\chi^{-1}$ is the inverse of the variance, given by:
\begin{equation}
\chi(|\mathbf r - \mathbf r'|) = \left<\delta\rho(\mathbf r)\delta\rho(\mathbf r')\right>=\bar\rho\delta(\mathbf r-\mathbf r')+\bar\rho^2[g(|\mathbf r-\mathbf r')-1]
\end{equation}
where $g(r)$ is the pair correlation function and $\bar\rho$ is the density of the fluid. 

We approximate the inverse as follows~\cite{Vaikuntanathan2014b,Chandler1993}
\begin{equation}
\int\int \chi^{-1} \approx \frac{1}{\int\int \chi}\,.
\end{equation}

To do so we need to compute:
\begin{equation}
\int d \mathbf r \int d \mathbf r' \chi(|\mathbf r - \mathbf r'|)= \int d \mathbf r \int d \mathbf r' (\bar\rho\delta(\mathbf r-\mathbf r')+\bar\rho^2[g(|\mathbf r-\mathbf r'|)-1]).
\end{equation}

The first term is:
\begin{equation}
 \int d \mathbf r \int d \mathbf r' \bar\rho\delta(\mathbf r-\mathbf r') =  \int d \mathbf r \bar\rho = V\bar\rho
\end{equation}

To find the second term, we first perform a change of variable. The correlation depends only on the magnitude $|\mathbf r-\mathbf r'|$. Let $\mathbf r'' = \mathbf r-\mathbf r'$, then:
\begin{align}
 \int d \mathbf r \int d \mathbf r' \bar\rho^2[g(|\mathbf r-\mathbf r'|)-1] &=  \int d \mathbf r' \int d \mathbf r'' \bar\rho^2[g(r'')-1]\\
 &= V\rho \int d \mathbf r \bar\rho[g(r)-1]
\end{align}
where in the last line we have renamed $r''$ as $r$.  

From the zero frequency limit of the Ornstein-Zernicke equation we have that~\cite{McQuarrie} 
\begin{equation}
1+\bar\rho \int d \mathbf r [g(r)-1] = \frac{1}{1-\bar\rho \int d \mathbf r c(r)}
\label{eq:oz}
\end{equation}
where $g(r)$ and $c(r)$ are the radial distribution function and direct pair correlation function, respectively. 
Using this:
\begin{equation}
\int d \mathbf r \int d \mathbf r' \chi(|\mathbf r - \mathbf r'|) = \frac{V\bar\rho}{1-\bar\rho \int d \mathbf r c(r)}\,.
\label{eq:loz}
\end{equation}

We can now compute the inverse, $\int\int \chi^{-1}$:
\begin{align}
\int\int \chi^{-1}&= \frac{1-\bar\rho \int d \mathbf r c(r)}{V\bar\rho} \\
& = \frac{1}{V\bar\rho}\int d \mathbf r\left[\delta(\mathbf r -\mathbf r')-\bar\rho c(r) \right]\\
&=  \int d \mathbf r \int d \mathbf r' \left[\frac{\delta(\mathbf r -\mathbf r')}{\bar\rho}-c(\mathbf r, \mathbf r')\right]
\end{align}

Substituting this back in to the Hamiltonian and taking the functional derivative with respect to $\delta\rho(\mathbf r)$ gives the effective force on $\delta\rho(\mathbf r)$:
\begin{equation}
\frac{\delta H}{\delta(\delta\rho)} = \beta \left[\frac{\delta\rho(\mathbf r)}{\bar\rho}-\int d \mathbf r' c(\mathbf r,\mathbf r')\delta\rho(\mathbf r')\right]
\end{equation}
 
Assuming that $\delta\rho$ is not spatially dependent, this looks like a linear restoring force:
\begin{equation}
 \frac{\delta H}{\beta \delta(\delta\rho)} \approx \delta\rho \left[\frac{1}{\bar\rho}-\int d \mathbf r c(r)\right] \equiv K_G \delta\rho = F_G
\label{eq:KG}
\end{equation}
where $K_G$ is the effective spring constant of the Gaussian density fluctuations. 
Further simplification can be achieved by using the Percus-Yevick closure~\cite{Chandler1993} to write down an approximate expression for $c(r)$, 
\begin{equation}
 K_G  =\left[\frac{1}{\bar\rho}-\int d \mathbf r c( r)\right] = \left[\frac{1}{\bar\rho}+\int d \mathbf r g(r)(e^{\beta V(r)} -1) \right]\,,
\label{eq:KG1}
\end{equation}
where $V(r)$ denotes the pairwise potential between particles. The term in Eq.~\ref{eq:KG1} proportional to $1/\bar\rho$ denotes an \textit{ideal gas} contribution and represents a long wavelength contribution.

\section{Proof $\langle {\mathbf F}^2 \rangle = \langle {\mathbf F}^2 \rangle _0 $ For a Liquid in Equilibrium}\label{sec:FsqFsqzero_eq}
%
%

Let us first evaluate $\langle {\bf F} ^2\rangle_0 = -\frac{\gamma}{2}  \displaystyle{\lim_{\Delta t \to 0}} \ex{{\bm \eta}(t) \cdot { \bf F}(t+\Delta t)}$. Assuming we have fixed a particle with position ${ \bf r}_1$ at the origin, by Taylor expanding the total force ${ \bf F}_1$ acting on this particle term by term we get the equation below in the $x$ direction and an analogous equation in the other directions:
\begin{align}
F_{1,x}(t+\Delta t) = F_{1,x}(t) + \sum_{i=1}^{N} \dfrac{\partial{F_{1,x}(t)}}{\partial{x_i}} \dot{x}_i(t) \Delta t + 1/2 \sum_{i=1}^{N} \sum_{j=1}^{N} \dfrac{\partial^2F_{1,x}(t)}{\partial x_i \partial x_j} \dot{x}_i(t) \dot{x}_j(t) (\Delta t)^2 + O(\Delta t^3)
\end{align}
Taking the dot product with the noise (in 2D to exemplify but easily generalizable to other dimensions) we obtain:
\begin{align}\label{Taylor}
 \displaystyle{\lim_{\Delta t \to 0}} \ex{{\bm \eta}_1(t) \cdot { \bf F}_{1}(t+\Delta t)} & = \displaystyle{\lim_{\Delta t \to 0}} \ex{( \eta_{1,x}(t), \eta_{1,y}(t)) \cdot (F_{1,x}(t+\Delta t), F_{1,y}(t+\Delta t) )} \\ 
& = \displaystyle{\lim_{\Delta t \to 0}} \mbox{\LARGE{(}} \ex{\eta_{1,x}(t) F_{1,x}(t) + \eta_{1,y}(t) F_{1,y}(t)} + \notag \\  & \sum_{i=1}^{N} \ex{\eta_{1,x}(t) \dfrac{\partial F_{1,x}(t)}{\partial x_i} \dot{x}_i(t) + \eta_{1,y}(t) \dfrac{\partial F_{1,y}(t)}{\partial y_i} \dot{y}_i(t) } \Delta t  + O(\Delta t^2)\mbox{\LARGE{)}} \label{Tayexp}
\end{align}
Terms of order $\Delta t^2$  or higher in Eq. \ref{Tayexp} vanish as $\Delta t$ goes to 0. Using the Langevin equation $\dot{x}_i(t) = \frac{1}{\gamma} F_{i,x}(t) + \eta_{i,x}(t)$, it follows that:

\begin{align}
\langle {\bf F} ^2\rangle_0 & = -\frac{\gamma}{2} \displaystyle{\lim_{\Delta t \to 0}} \ex{{\bm \eta}(t) \cdot { \bf F}(t+\Delta t)} \nonumber \\
& = -\frac{\gamma}{2} \displaystyle{\lim_{\Delta t \to 0}} \ex{\dfrac{\partial F_{1,x}(t)}{\partial x_1} \eta^2_{1,x}(t) +  \dfrac{\partial F_{1,y}(t)}{\partial y_1}  \eta^2_{1,y}(t)} \Delta t \\
& = \frac{\gamma}{2}\displaystyle{\lim_{\Delta t \to 0}} \left( \ex{\dfrac{\partial^2 U}{\partial x_1^2}} \dfrac{\ex{dW_x^2}}{\Delta t} + \ex{\dfrac{\partial^2 U}{\partial y^2_1}} \dfrac{\ex{dW_y^2}}{\Delta t} \right) \\
& = K_B T  \ex{\grad^2_{1} U} \\
& =  K_B T \rho \int  \grad^2 u({\bf r})  g({\bf r}) d{\bf r} \label{fsqzerofinal}
\end{align}

In the above, we have used the fact that $ \langle dW_{x}^2 \rangle = \langle (\eta_x \Delta t)^2 \rangle= 2 D_0 \Delta t = \frac{2K_B T}{\gamma} \Delta t $. We also wrote the total potential energy $U$ as a sum of pairwise potentials $u({ \bf r} _{ij})$ and replaced the summation with an integral over the pair correlation function, $g({\bf r})$.

Let us now expand the term $\ex{{\mathbf F}^2}$. Again by focusing on a single particle with position ${ \bf r}_1$ fixed at the origin and doing similar manipulations we obtain:
\begin{align}
\ex{{\bf F}^2} & = \ex{(\grad_{1} U({\bf r}))^2} \nonumber \\
& =  \rho \int (\grad u({\bf r}))^2 g({\bf r})  d{\bf r}  \notag \\
& + \rho^2 \int  \int  \grad u({\bf r}) \cdot  \grad u({\bf r}')  g_3({\bf r}, {\bf r}') d{\bf r} d{\bf r}' \label{Fsq}
\end{align}

To aid us in evaluating Eq. \ref{Fsq}, we assume the system is in equilibrium with a Boltzmann distributed steady state, and use the Yvon-Born-Green hierarchy: ~\cite{rowlinson2013}
\begin{align}
 g({\bf r})\grad u({\bf r}) = -K_B T \grad g({\bf r}) -  \rho \int d{\bf r}' \grad u({\bf r}') g_3({\bf r}, {\bf r}')
\end{align}
Plugging this expression into Eq. \ref{Fsq} cancels the term containing the three-point correlation function, yielding:
\begin{align}
\ex{{\bf F}^2} & = -K_B T \rho \int  \grad u({\bf r}) \cdot  \grad g({\bf r}) d{\bf r} \label{Fsqfirst} 
\end{align}
Integrating by parts and setting boundary terms to zero gives the final expression for $\langle {\bf F}^2  \rangle$:
\begin{align}
\ex{{\bf F}^2} & =   K_B T \rho \int   \grad^2 u({\bf r})  g({\bf r}) d{\bf r} 
\end{align}  
The expression is equal to that in Eq. \ref{fsqzerofinal}, completing the proof that $\langle {\mathbf F}^2 \rangle = \langle {\mathbf F}^2 \rangle _0 $ at equilibrium.

\section{Expression for Diffusion from Perturbation Theory}\label{sec:PerturbationTheory}
%
%

The expression for diffusion in terms of force-force and noise-force correlations is: 
\begin{equation}
D-D_0= \frac{1}{d} \int_0^\infty  \left[\left<\frac{{\bf F}(0) \cdot {\bf F}(t)}{\gamma^2}\right>+\left<\frac{{\boldsymbol \eta}(0) \cdot {\bf F}(t)}{\gamma}\right>\right]{d}t 
\label{eq:KGD1}
\end{equation}
To set up our perturbative approach to computing the quantities in Eq. \ref{eq:KGD1}, we start from the overdamped Langevin equation:
\begin{equation}
 \dot{{\bf r}} = \frac{h}{\gamma}{\bf \tilde F} ({\bf r})  + \tilde {\bm\eta}(t) =  \frac{h}{\gamma} \left( {\bf \tilde F}_c ({\bf r}) +  Pe {\bf \tilde F}_d ({\bf r})\right) +  \tilde {\bm\eta} (t)\label{Langevinpert}
\end{equation}
where ${\bf \tilde F}_c ({\bf r}) = -\nabla \tilde U({\bf r})$ is the conservative force and ${\bf \tilde F}_d ({\bf r}) = {\bf \nabla} \times {\bf \tilde A}( {\bf r})$ the dissipative force. ${\tilde U}({\bf r})$ and ${\bf \tilde A}({\bf r})$ obey $\int  {\tilde U}({\bf r}) d{\bf r} = 0$ and $\nabla \cdot {\bf \tilde A}({\bf r}) = 0$. We will be working in the limit $Pe\,,h << 1$. The probability distribution for this system obeys the Fokker-Planck equation:

\begin{equation}
 \frac{\partial P}{\partial t} = - h \vec{\nabla} \cdot \left [{\bf \tilde F}_c P \right] + D \nabla^2 P - h Pe \vec{\nabla} \cdot \left[  {\bf \tilde F}_d P \right] 	.\label{FokkerPlanckpert}
\end{equation}

When $Pe=0$, the system is in equilibrium and Eq.\,\ref{FokkerPlanckpert} is solved simply by the Boltzmann distribution $P_0 = \dfrac{e^{-\beta h U({\bf r})}}{Z}$, where $Z$ is the partition function. We construct a perturbation theory in $Pe$ by considering the latter term in Eq. \ref{FokkerPlanckpert} as a perturbation to the equilibrium distribution. The first correction to the equilibrium distribution is then given by:

\begin{equation}
 P_1 = Pe \sum_i \frac{ | u_i \rangle \langle v_i | {\bf \tilde F}_d \cdot \nabla  |P_0  \rangle } {\lambda_i}	\label{P1}
\end{equation}
where the vectors $| u_i \rangle $ and $ \langle v_i |$  are the right and left eigenvectors of $W_0$, respectively, and $\lambda_i$ are the eigenvalues.
The total probability for our system is then given by:

\begin{equation}
 P({\bf r}) = \dfrac{e^{-\beta h U({\bf r})}}{Z} + h Pe \sum_i \frac{ | u_i \rangle \langle v_i | {\bf \tilde F}_d  \cdot \nabla  |P_0  \rangle } {\lambda_i}	\label{P}
\end{equation}

To proceed with evaluating the expression for the diffusion coefficient perturbatively. Defining $\epsilon_1\equiv h$ and $\epsilon_2\equiv Peh$, we will evaluate terms to second order in $\epsilon_{1,2}$. In order for mixed third order terms such as $\epsilon_1^2 \epsilon_2$ to be smaller than all quadratic  we need $h < Pe$ in our setup. We will average the force-force correlations and the noise-force correlations only over the noise in the supplementary information. The averaging over the realizations of the force are implied at the end and shown as angular brackets in the main text. We first consider the expression for $\int_0^{\infty}  \langle  {\bf \tilde F}(0) \cdot  {\bf \tilde F}(t)  \rangle dt$:

\begin{align}
h^2 \int_0^{\infty}  \langle  {\bf \tilde F}(0) \cdot  {\bf \tilde F}(t)  \rangle dt & = h^2 \int_0^{\infty} dt \int d{\bf r} \frac{ P({\bf r})}{Z} \int \dfrac{d {\bf q}}{(2\pi)^d} \int \dfrac{d  {\bf q}'}{(2\pi)^d}  {\bf \tilde F}( {\bf q}) \cdot  {\bf \tilde F}( {\bf q}') e^{i( {\bf q}+ {\bf q}') \cdot{\bf r}} \langle   e^{\int_0^{t} i {\bf q}' \cdot {\bm \tilde \eta}(t') dt' + h/\gamma \int_0^{t}   i {\bf  {\bf q}'} \cdot {\bf \tilde F}(t') dt' } \rangle
\end{align}
Keeping terms to order $\epsilon_{1,2}^2$ we have:
\begin{align}
h^2 \int_0^{\infty}  \langle  {\bf \tilde F}(0) \cdot  {\bf \tilde F}(t)  \rangle dt & \approx h^2 \frac{1}{V} \int \dfrac{d  {\bf q}}{(2\pi)^d} \int \dfrac{d  {\bf q}'}{(2\pi)^d}  {\bf \tilde F}( {\bf q})   \cdot {\bf \tilde F} ({\bf q}') \delta( {\bf q} + {\bf q}' )  \int_0^{\infty} \langle e^{ \int_0^t  i{\bf q}' \cdot {\bm \tilde \eta}(t') dt'} \rangle  dt \\
 & = h^2\frac{1}{V} \int \dfrac{d  {\bf q}}{(2\pi)^d}  {\bf \tilde F}( {\bf q}) \cdot  {\bf \tilde F}(- {\bf q}) \int_0^{\infty}  e^{-|{\bf q}|^2 D_0 t} dt \\
 & = h^2 \frac{1}{D_0V} \int \dfrac{d  {\bf q}}{(2\pi)^d} \frac{1}{| {\bf q}|^2} {\bf \tilde F}( {\bf q}) \cdot    {\bf \tilde F}( {\bf -q})  \\
& =  h^2 \frac{1}{D_0V} \int \dfrac{d  {\bf q}}{(2\pi)^d} \frac{1}{| {\bf q}|^2} \left({\bf \tilde F}_c( {\bf q}) \cdot    {\bf \tilde F}_c( {\bf -q})  + 2 Pe {\bf \tilde F}_c( {\bf q}) \cdot    {\bf \tilde F}_d( {\bf -q}) + Pe^2 {\bf \tilde F}_d( {\bf q}) \cdot    {\bf \tilde F}_d( {\bf -q}) \right) \label{FF}
\end{align}

The noise-force correlation term is given by:

\begin{align}
h \int_0^{\infty} dt \langle  \tilde {\bm\eta}(0) \cdot  {\bf \tilde F}(t)  \rangle  = h \int_0^{\infty} dt \int d{\bf r} \frac{ P({\bf r})}{Z} \int \dfrac{d {\bf q}}{(2\pi)^d} \langle \tilde {\bm\eta}(0) \cdot {\bf \tilde F}( {\bf q})  e^{i{\bf q} \cdot{\bf r}}    e^{\int_0^{t} i {\bf q} \cdot  \tilde {\bm\eta}(t') dt' + h/\gamma \int_0^{t}   i {\bf  {\bf q}} \cdot {\bf \tilde F}(t') dt' } \rangle  \label{etaF}
\end{align}

The first term of order $\epsilon_{1,2}^2$, which we call $T_1$, comes from $P({\bf r})$, which we expand in terms of $h$: 

\begin{align}
T_1 &= \frac{h^2}{V} \int d{\bf r}( -\beta {\tilde U}({\bf r})) \int \dfrac{d {\bf q}}{(2\pi)^d} e^{i{\bf q} \cdot{\bf r}}   \int_0^{\infty} \langle  \tilde {\bm\eta}(0) \cdot {\bf \tilde F}( {\bf q}) e^{\int_0^{t} i {\bf q} \cdot  \tilde {\bm\eta}(t') dt'} \rangle dt \\
& = \frac{-\beta h^2}{V}   \int \dfrac{d {\bf q}}{(2\pi)^d} \int \dfrac{d {\bf q}'}{(2\pi)^d} {\tilde U}({\bf q'}) \delta{ ({\bf q} + {\bf q}'})  \int_0^{\infty} \langle \tilde {\bm\eta}(0) \cdot {\bf \tilde F}( {\bf q})  e^{\int_0^{\Delta t} i {\bf q} \cdot \tilde {\bm\eta}(t') dt'}  e^{\int_{\Delta t}^{t} i {\bf q} \cdot \tilde {\bm\eta}(t') dt'} \rangle dt \\
& = \frac{-2 D_0 \beta h^2}{V}   \int \dfrac{d {\bf q}}{(2\pi)^d} {\tilde U}({\bf -q})   i {\bf q} \cdot {\bf \tilde F}( {\bf q})  \int_0^{\infty} e^{-|{\bf q}|^2 D_0 t} dt\\ 
& = \frac{-2 \beta h^2}{V}   \int \dfrac{d {\bf q}}{(2\pi)^d} \frac{1}{|{\bf q}|^2} \left ( {\bf \tilde F}_c({ - \bf q}) \right) \cdot {\bf \tilde F}( {\bf q}) \\ 
& = -\frac{2 \beta h^2}{V}   \int \dfrac{d {\bf q}}{(2\pi)^d} \frac{1}{|\bf q|^2} \left( {\bf \tilde F}_c( {\bf -q}) \cdot {\bf \tilde F}_c( {\bf q}) + Pe {\bf \tilde F}_c( {\bf -q}) \cdot {\bf \tilde F}_d( {\bf q}) \right) \label{etaFfinal}
\end{align}

The second term quadratic in $\epsilon_{1,2}$, $T_2$, comes from the force present in the exponential in Eq. \ref{etaF}. Ignoring constants, this term can be written as:

\begin{align}
T_2 & \propto h^2 \int d{\bf r} \int \dfrac{d {\bf q}}{(2\pi)^d} \int \dfrac{d {\bf q}'}{(2\pi)^d}   e^{i( {\bf q} + {\bf q}') \cdot {\bf r}} i {\bf q} \cdot {\bf \tilde F}( {\bf q}') \int_0^{\infty} dt \int_0^t dt' \langle {\bf \tilde F}( {\bf q}) \cdot \tilde {\bm\eta}(0) e^{\int_0^{t} i {\bf q} \cdot  \tilde {\bm\eta}(\tau) d\tau} e^{\int_0^{t'} i {\bf q'} \cdot \tilde {\bm\eta}(\tau) d\tau} \rangle \\ 
& = h^2 \int \dfrac{d {\bf q}}{(2\pi)^d}  i {\bf q} \cdot {\bf \tilde F}( {\bf -q}) \int_0^{\infty} dt \int_0^t dt' \langle {\bf \tilde F}( {\bf q}) \cdot \tilde {\bm\eta}(0) e^{\int_{t'}^{t} i {\bf q} \cdot \tilde {\bm\eta}(\tau) d\tau} \rangle \\
& = h^2 \lim_{\Delta t \to 0} \int \dfrac{d {\bf q}}{(2\pi)^d} i {\bf q} \cdot {\bf \tilde F}( {\bf -q}) \int_0^{\infty} dt \int_0^{\Delta t}  dt' \langle {\bf \tilde F}({\bf q}) \cdot \tilde {\bm\eta}(0) e^{\int_{t'}^{t} i {\bf q} \cdot \tilde {\bm\eta}(\tau) d\tau} \rangle  \\ 
& = h^2 \lim_{\Delta t \to 0} \int  \dfrac{d {\bf q}}{(2\pi)^d}  i {\bf q} \cdot {\bf \tilde F}( {\bf -q}) \int_0^{\infty} dt \langle {\bf \tilde F}({\bf q}) \cdot \tilde {\bm\eta}(0) e^{\int_{0}^{t} i {\bf q} \cdot \tilde {\bm\eta}(\tau) d\tau}  \rangle \Delta t  \\
& = 2 h^2 \lim_{\Delta t \to 0} \int  \dfrac{d {\bf q}}{(2\pi)^d} i {\bf q} \cdot {\bf \tilde F}( {\bf -q})  i {\bf q} \cdot {\bf \tilde F}( {\bf q}) \frac{1}{|{\bf q}|^2}  \Delta t = 0
\end{align}

The term linear in $h$, let's call it $T_3$, can be computed in a similar way, resulting in the following expression:

\begin{align}
T_3 & \propto  h \int_0^{\infty} dt \int d{\bf r} \int \dfrac{d {\bf q}}{(2\pi)^d} \langle \tilde {\bm\eta}(0) \cdot {\bf \tilde F}( {\bf q})  e^{i{\bf q} \cdot{\bf r}}    e^{\int_0^{t} i {\bf q} \cdot \tilde {\bm\eta}(t') dt'} \rangle \\
 & = 2 h\int d{\bf r} \int \dfrac{d {\bf q}}{(2\pi)^d} \frac{1}{|\bf q|^2} e^{i{\bf q} \cdot {\bf r}} i{\bf q} \cdot \left( {\bf \tilde F}_c( {\bf q}) + Pe {\bf \tilde F_d}( {\bf q}) \right)  \\
 & = 2 h \left( \int d{\bf r} { \tilde U}({\bf r}) +  Pe \int \dfrac{d {\bf q}}{(2\pi)^d} \frac{1}{|\bf q|^2}  i{\bf q} \cdot  {\bf \tilde F_d}( {\bf q})  \right) \\
 & = 2 h \left( \int d{\bf r} { \tilde U}({\bf r}) +  Pe \int \dfrac{d {\bf q}}{(2\pi)^d} \frac{1}{|\bf q|^2}  i{\bf q} \cdot i{\bf q} \times {\bf \tilde A} ({\bf q}) \right) \\ 
& = 2 h \left( \int d{\bf r} {\tilde U}({\bf r}) +  Pe \int \dfrac{d {\bf q}}{(2\pi)^d} \frac{1}{|\bf q|^2}  i{\bf q} \times i{\bf q} \cdot {\bf \tilde A}({\bf q})  \right) = 0
\end{align}

Plugging Eq. \ref{FF} and Eq. \ref{etaFfinal} into the expression \ref{eq:KGD1}, and using ${\bf \tilde F_c}({\bf  q}) = -i{\bf q} \tilde U({\bf q})$ and ${\bf \tilde F_d}({\bf  q}) = i{\bf q}  \times {\bf \tilde A}({\bf q})$ yields the desired relationship:

\begin{align}
D-D_0= \frac{D_0 h^2}{(K_B T)^2dV}\left(Pe^2 \int  |{\bf \tilde A}({\bf r})|^2 d{\bf r} - \int   \tilde U^2({\bf r})d{\bf r} \right) \label{diffusion}
\end{align}

We can also use this perturbation theory to obtain expressions for the quantities $\langle { \bf \tilde F}^2 \rangle$ and $\langle {\bf \tilde F}^2 \rangle_0$:

\begin{align}
\langle {\bf \tilde F}^2 \rangle & = \frac{h^2}{V} \int d{\bf r}  \int \dfrac{d {\bf q}}{(2\pi)^d} \int \dfrac{d {\bf q}'}{(2\pi)^d}     {\bf \tilde F}( {\bf q}) \cdot  {\bf \tilde F}( {\bf q}') e^{i( {\bf q} + {\bf q}')} \\ 
& =  \frac{h^2}{V} \int \dfrac{d {\bf q}}{(2\pi)^d} \left(   {\bf \tilde F}_c( {\bf q}) \cdot  {\bf \tilde F}_c( {\bf -q}) + 2Pe{\bf \tilde F}_d( {\bf q}) \cdot  {\bf \tilde F_c}( {\bf -q}) + Pe^2{\bf \tilde F}_d( {\bf q}) \cdot  {\bf \tilde F}_d( {\bf -q}) \right) \label{Fsq_diss}
\end{align}

\begin{align}
\langle {\bf \tilde F}^2\rangle_0  & = \frac{h^2}{V} \frac{\beta\gamma}{2}  \int d{\bf r} \tilde U({\bf r}) \int \dfrac{d {\bf q}}{(2\pi)^d}  e^{i{\bf q} \cdot {\bf r}}  \langle e^{i {\bf q} \cdot  \tilde {\bm\eta}(0) \Delta t}  \tilde {\bm\eta}(0) \cdot {\bf \tilde F}( {\bf q}) \rangle  \\ 
& =  \frac{h^2}{V} \beta\gamma D_0  \int \dfrac{d {\bf q}}{(2\pi)^d} \int \dfrac{d {\bf q}'}{(2\pi)^d}  e^{i({\bf q} + {\bf q}')\cdot {\bf r}}  i{\bf \tilde q} \cdot {\bf \tilde F}( {\bf q}) {\tilde U}({\bf q}') \\
& =  \frac{ h^2}{V} \int \dfrac{d {\bf q}}{(2\pi)^d}  {\bf \tilde F}( {\bf q}) \cdot {\bf \tilde F}_c( {\bf -q}) \\
& = \frac{ h^2}{V}  \int \dfrac{d {\bf q}}{(2\pi)^d}  \left( {\bf \tilde F}_c( {\bf q}) \cdot {\bf \tilde F}_c( {\bf -q}) + Pe{\bf \tilde F}_d( {\bf q}) \cdot {\bf \tilde F}_c( {\bf -q}) \right) \label{Fsq0_diss}
\end{align}

Subtracting Eq. \ref{Fsq0_diss} from Eq.  \ref{Fsq_diss} yields the relationship for entropy dissipation: 

\begin{align}
\langle {\bf \tilde F}^2 \rangle - \langle {\bf \tilde F}^2\rangle_0  & = 
 \frac{ h^2}{V}  \int \dfrac{d {\bf q}}{(2\pi)^d}   \left( Pe{\bf \tilde F}_c( {\bf q}) \cdot {\bf \tilde F}_d( {\bf -q}) + Pe^2 {\bf \tilde F}_d( {\bf q}) \cdot {\bf \tilde F}_d( {\bf -q}) \right) \\
 & = \frac{ h^2}{V} Pe^2 \int \dfrac{d {\bf q}}{(2\pi)^d}  {\bf \tilde F}_d( {\bf q}) \cdot {\bf \tilde F}_d( {\bf -q}) 
\end{align}

In the above we used the fact that $\int \dfrac{d {\bf q}}{(2\pi)^d} {\bf \tilde F}_c( {\bf q}) \cdot {\bf \tilde F}_d({\bf -q})$ can be rewritten as  $\int d {\bf r} {\bf \tilde F}_c ({\bf r}) \cdot \curl {\bf \tilde A ({\bf r})} = - \int d {\bf r}  {\bf \tilde A ({\bf r})} \cdot  \curl {\bf \tilde F}_c ({\bf r}) = 0 $. When $Pe=0$, entropy production is zero and $\langle {\bf \tilde F}^2 \rangle = \langle {\bf \tilde F}^2 \rangle_0$.

We simulated a particle in a 2D periodic force landscape $F = (- h \cos(x) + hPe \cos(y), -h \cos(y) - hPe cos(x) )$. We estimated $D_{eq}$ in such a system by running 5 simulations with $Pe = 0$ and averaging the resulting diffusion constants.  Then we varied $Pe$ between 0.2 and 0.45, and computed the diffusion constant and the error in each case by averaging over 5 simulations. Next we report how the deviation of the diffusion constant from $D_{eq}$ in our simulations compares to the expected correction due to $Pe$, which is $h^2Pe^2/2$. For this specific system, we set ${\bf F}_d \cdot {\bf F}_c = 0$ thus ensuring that $P_1=0$ (Eq.~\ref{P1}). Hence, although $h > Pe$ in these simulations, the correction to the diffusion constant arising from the nonequilibrium forces will scale like $Pe^2$ to leading order. 

\begin{figure}
\centering
\includegraphics[width=0.9\linewidth]{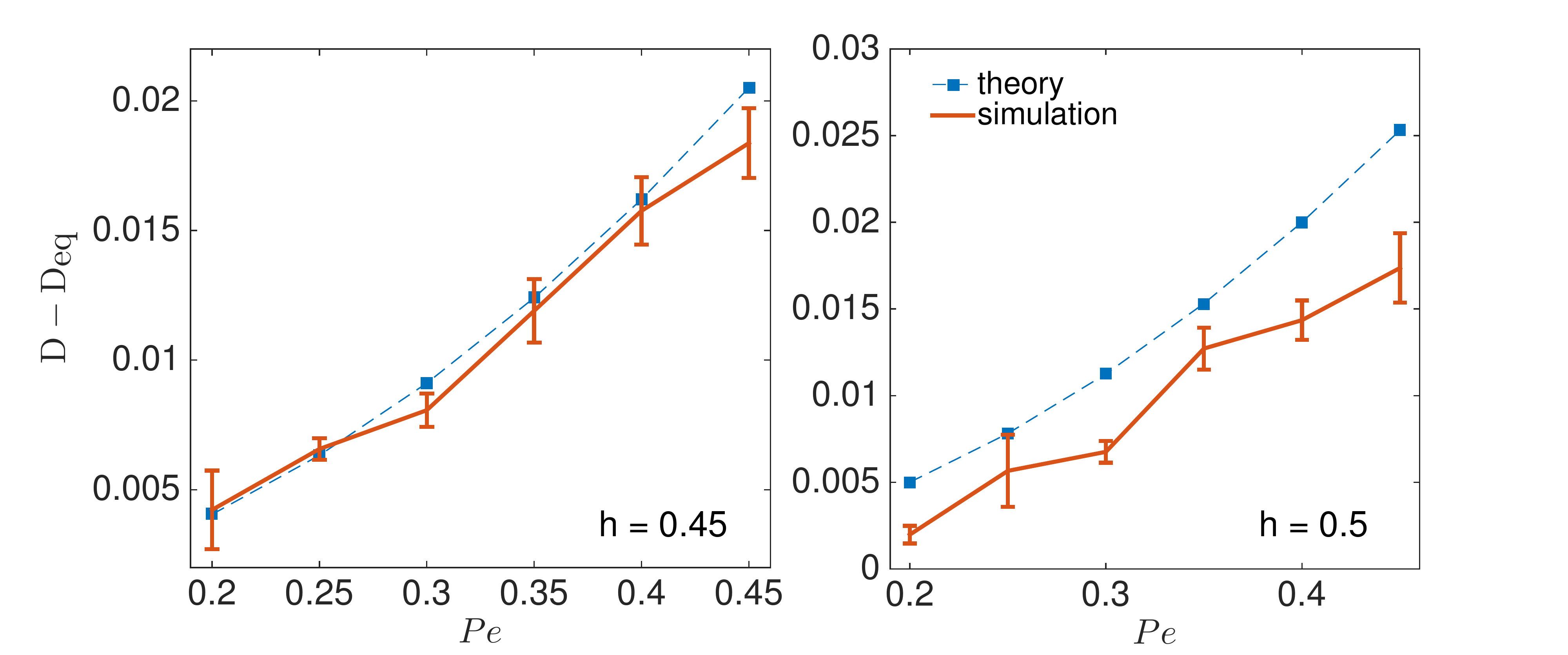}
\caption{The correction to the diffusion constant due to the nonconservative force is plotted as a function of $Pe$ for a system characterized by a periodic force landscape. $D_{eq}$ is computed as the average diffusion constant for the system with $Pe$ = 0. The data points are reasonably within error bars for $h = 0.45$, while there is some systematic deviation at $h = 0.5$.}
\label{fig:diffusion2D}
\end{figure}

\section{Comparing our Diffusion Formula to other Analytical Expressions at Equilibrium} \label{sec:Comparison}

At equilibrium, our prediction for the  diffusion coefficient is given by:
\begin{equation}
D/D_0 = 1 - \frac{h^2}{(K_B T)^2dV} \int  U^2({\bf r}) d {\bf r}
\end{equation}
Here we compare our prediction to existing ones in the literature. In 1D, the analytical expression for the diffusion of a tagged particle in a periodic potential is due to ~\cite{lifson1962} and reads:  $D/D_0 = 1/ \left( \langle e^{-\beta U} \rangle  \langle e^{\beta U} \rangle \right)$, where the expectation value is taken to mean $\langle f(x) \rangle = 1/L \int_0^L f(x) dx$. In our perturbation theory, this expression becomes, to order $h^2$:
\begin{align}
\dfrac{D}{D_0} & = \dfrac{1}{\int_0^L e^{-\beta h U(x)} dx \int_0^L e^{\beta h U(x')} dx'} \\
& = \dfrac{L^2}{( L - h \beta \int_0^L U(x) dx  + (h^2\beta^2/2)  \int_0^L U^2(x) dx + \dots)( L + h \beta \int_0^L U(x') dx'  + (h^2\beta^2/2)  \int_0^L U^2(x') dx' + \dots)} \\
& \approx \dfrac{1}{ 1 + \frac{h^2\beta^2}{L} \int_0^L U^2(x) dx} \\
& \approx 1 - \frac{h^2\beta^2}{L} \int_0^L U^2(x) dx
\end{align}
which is in agreement with our results. In 2D the corresponding formula $D/D_0 = 1/  \langle e^{\beta U} \rangle$  is due to ~\cite{dean2007}. Expanding this expression to order $h^2$ and  recognizing that $\int_0^L \int_0^L U(x,y) dx dy =0$, we get:
\begin{align}
\dfrac{D}{D_0} & = \dfrac{L^2}{\int_0^L \int_0^L e^{\beta h U(x,y)} dx dy} \\
& = \dfrac{L^2}{\int_0^L \int_0^L \left( 1 + h \beta U(x,y) + (h^2\beta^2/2) U^2(x,y) +\dots \right) dx dy} \\
& \approx \dfrac{1}{ 1 + \frac{h^2\beta^2}{2L^2} \int_0^L \int_0^L  U^2(x,y) dx dy } \\
& \approx 1 - \frac{h^2\beta^2}{2L^2} \int_0^L \int_0^L  U^2(x,y) dx dy  
\end{align}
This result is also in agreement with our prediction.

\section{Higher Order Terms in Expression for Diffusion} \label{sec:HigherOrderTerms}

We now show that higher order terms in $h$ do not diverge. Higher order terms of $\int_0^{\infty} {\bf \tilde F}(0) \cdot {\bf \tilde F}(t) dt$ can be expressed generally as:
\begin{align}
H_{nm} & \propto \frac{h^n}{(D_0)^{m-1} (2\pi)^{d(n-1)}} \int d {\bf q}_2 \dotso \int d{\bf q}_n {\bf \tilde F}\left(-{\textstyle \sum\limits_{i=2}^n{\bf q}_i} \right) \cdot {\bf \tilde F}({\bf q}_2) \frac{ i{\bf q}_2 \cdot {\bf \tilde F}({\bf q}_3) \dotso i{\bf q}_{m-1} \cdot {\bf \tilde F}({\bf q}_m) {\tilde U}({\bf q}_{m+1}) \dotso {\tilde U}({\bf q}_{n})}{|{\bf q}_2|^2 ( |{\bf q}_2|^2 + |{\bf q}_3 |^2) \dotso ( |{\bf q}_2|^2 + |{\bf q}_3 |^2 + \dotso + |{\bf q}_m |^2)}
\end{align}
where $n \geq 3$ and $n \geq m \geq 2$. 
This quantity scales with system size as $L^{(1-n)d+(2-m)+2(m-1)}$. When $m=n$, we get a maximum scaling with system size as $L^{(1-n)d+n}$. For $d=1$, higher order terms in $n$ can diverge with system size. For $d=2$, the scaling is $L^{2-n}$, which can diverge only for terms of order less than or equal to 2. (Therefore, as long as $\langle F \rangle =0$ holds, then the quadratic terms are well behaved as described in Sec.\ref{sec:PerturbationTheory}). For d=3, the scaling is $L^{3-2n}$, which can not diverge since $n \geq 2$.

The higher order terms in $\int_0^{\infty} {\bm\eta}(0) \cdot {\bf \tilde F}(t) dt$ can also be conveniently expressed as:
\begin{align}
H_{nm} & \propto \frac{2 h^n}{(D_0)^{m-1} (2\pi)^{d(n-1)}} \int d {\bf q}_1 \dotso \int d{\bf q}_{n-1} i{\bf q}_1 \cdot {\bf \tilde F}({\bf q}_1) \frac{ i{\bf q}_1 \cdot {\bf \tilde F}({\bf q}_2) \dotso i{\bf q}_{m-1} \cdot {\bf \tilde F}({\bf q}_m) {\tilde U}({\bf q}_{m+1}) \dotso {\tilde U}\left(-{\textstyle \sum\limits_{i=1}^{n-1}}{\bf q}_i \right)}{|{\bf q}_1|^2 \left( |{\bf q}_1|^2 + |{\bf q}_2 |^2 \right) \dotso ( |{\bf q}_1|^2 + |{\bf q}_2 |^2 + \dotso + |{\bf q}_m |^2)}
\end{align}
where $n \geq 3$ and $n-1 \geq m \geq 1$. Terms with $m=n$ in this case are zero.

This quantity scales with system size as $L^{(1-n)d+(-m)+2m}$. When $m=n$, we obtain the maximum scaling with system size but these terms are null so the higher order terms cannot diverge.

\section{Scaling Relations in Other Kinds of Driving}\label{sec:other_driving}
\begin{figure}[!htb]
\minipage{0.32\textwidth}
  \includegraphics[width=\linewidth]{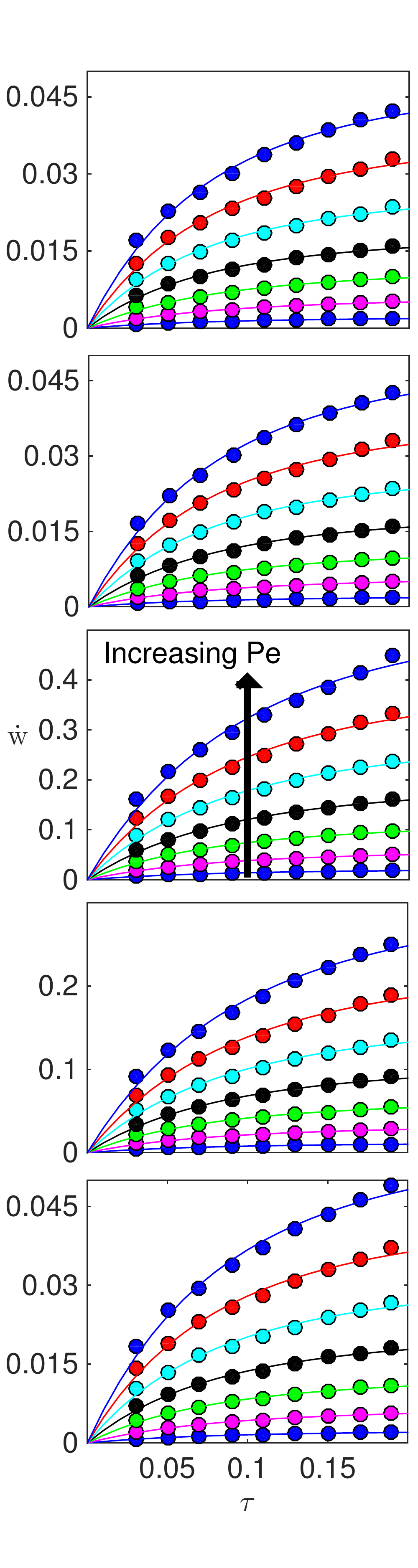}
\endminipage\hfill
\minipage{0.32\textwidth}
  \includegraphics[width=\linewidth]{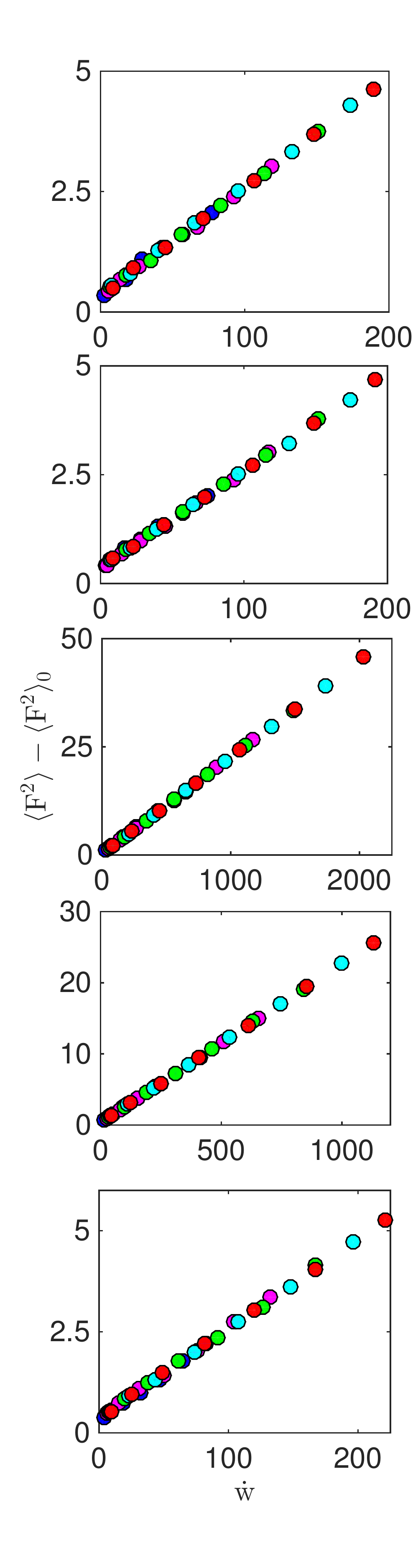}
\endminipage\hfill
\minipage{0.32\textwidth}
  \includegraphics[width=\linewidth]{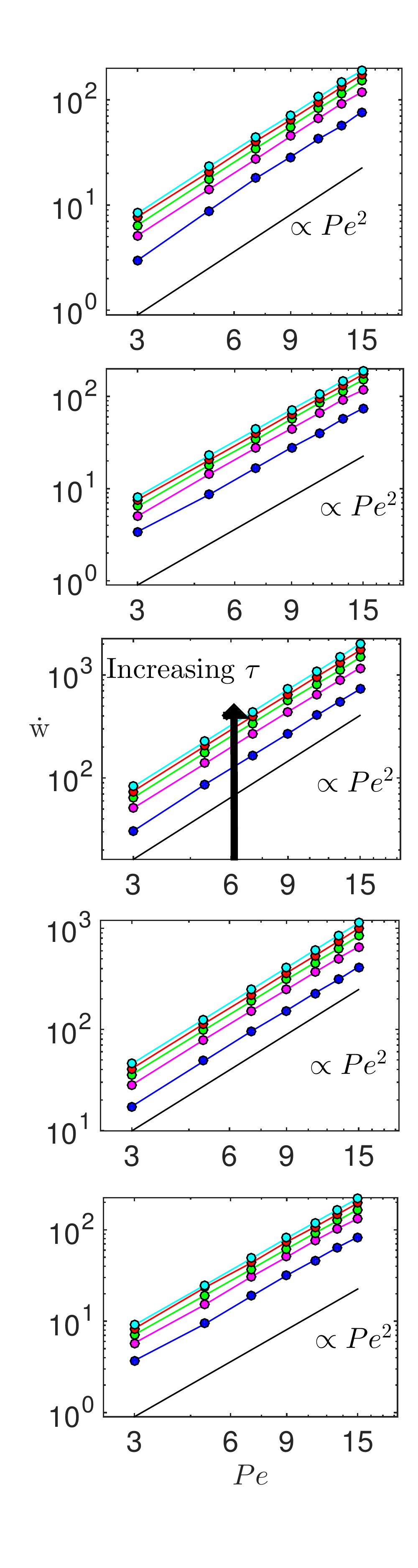}
\endminipage
\caption{We show the scaling of the work with $\tau$ (left), force fluctuations (center) and $Pe^2$ (right). Top two rows correspond to in-phase driving like in the main text, but the fraction of active particles is, in this order, 0.9 and 0.1. The bottom three rows correspond to active particles with initial phases drawn from a uniform distribution from 0 to $2 \pi$ and fractions of active particles 0.9, 0.5 and 0.1, in this order. Error bars are smaller than the points. In the scaling of $\langle F \rangle^2 - \langle F^2 \rangle_0$ with the work, the $\tau$ values sampled were 3, 7, 11, 15 and 19. In the scaling with $\tau$, the $K_G$ and $\gamma_G$ parameters can vary between the five investigated systems, but for each system the variation in these parameters is small. Maximum variation in $K_G$ was $\sim 14 \%$ and maximum variation in $\gamma_G$ was $\sim 11\%$. } \label{fig:other_driving}
\end{figure}

In Fig.~\ref{fig:other_driving}, we verify the scaling of work, force fluctuations for other kinds of driving forces. This establishes their general nature.


%
